\documentclass[%
 reprint,
superscriptaddress,
 amsmath,amssymb,
 aps,floatfix
]{revtex4-2}

\usepackage[pdftex]{graphicx}% Include figure files
\usepackage{dcolumn}% Align table columns on decimal point
\usepackage{bm}% bold math
\usepackage{xfrac}
\usepackage{booktabs}
\usepackage[table,xcdraw]{xcolor}
\newcolumntype{L}{>{$}l<{$}}
\newcolumntype{C}{>{$}c<{$}}
\newcolumntype{R}{>{$}r<{$}}

\begin{document}

\preprint{APS/123-QED}

\title{Curvature directed anchoring and defect structure of colloidal smectic liquid crystals in confinement}

\author{Ethan I. L. Jull}
 \email{e.i.l.jull@uu.nl}
 \affiliation{Soft Condensed Matter and Biophysics, Debye Institute for Nanomaterials Science, Utrecht University, 3584 CC Utrecht, The Netherlands}
\author{Gerardo Campos-Villalobos}
 \email{g.d.j.camposvillalobos@uu.nl}
 \affiliation{Soft Condensed Matter and Biophysics, Debye Institute for Nanomaterials Science, Utrecht University, 3584 CC Utrecht, The Netherlands}
\author{Qianjing Tang}
% \email{q.tang@uu.nl}
 \affiliation{Department of Earth Sciences, Utrecht University, 3584 CB Utrecht, The Netherlands}
\author{Marjolein Dijkstra}
 \email{m.dijkstra@uu.nl}
 \affiliation{Soft Condensed Matter and Biophysics, Debye Institute for Nanomaterials Science, Utrecht University, 3584 CC Utrecht, The Netherlands}
\author{Lisa Tran}
 \email{l.tran@uu.nl}
 \affiliation{Soft Condensed Matter and Biophysics, Debye Institute for Nanomaterials Science, Utrecht University, 3584 CC Utrecht, The Netherlands}

\date{\today}

\begin{abstract}
Rod-like objects at high packing fractions can form smectic phases, where the rods break rotational and translational symmetry by forming lamellae. Smectic defects thereby include both discontinuities in the rod orientational order (disclinations), as well as in the positional order (dislocations). In this work, we use both experiments and simulations to probe how local and global geometrical frustrations affect defect formation in hard-rod smectics. We confine a particle-resolved, colloidal smectic within elliptical wells of varying size and shape for a smooth variation of the boundary curvature. We find that the rod orientation near a boundary --- the anchoring --- depends upon the boundary curvature, with an anchoring transition observed at a critical radius of curvature approximately twice the rod length. The anchoring controls the smectic defect structure. By analyzing local and global order parameters, and the topological charges and loops of networks made of the density maxima (rod centers) and density minima (rod ends), we quantify the amount of disclinations and dislocations formed with varying confinement geometry. More circular confinements, having only planar anchoring, promote disclinations, while more elliptical confinements, with antipodal regions of homeotropic anchoring, promote long-range smectic ordering and dislocation formation. Our findings demonstrate how geometrical constraints can control the anchoring and defect structures of liquid crystals - a principle that is applicable from molecular to colloidal length scales.
%\begin{description}
%\end{description}
\end{abstract}

\maketitle

\section{\label{sec:Intro}Introduction}

Topological defects are pervasive in diverse phases of ordered matter, and their presence greatly affects system properties \cite{ChaikinLubensky1995}. Liquid crystals are prime examples \cite{deGennes}, where topological defects facilitate optical applications \cite{KhooWu1993}, particle assembly \cite{Musevic2006,Hegmann2007,Tran2020}, and shape transformations in elastomers \cite{WarnerTerentjev2007}. At even larger length scales, liquid crystal defects influence biological processes in bacterial biofilm development \cite{DellArciprete2018,Yaman2019,Copenhagen2021,Prasad2023}, cell proliferation, and morphogenesis \cite{Doostmohammadi2016,Kawaguchi2017,Saw2017,Maroudas-Sacks2021,VafaMahadevan2022,Kaiyrbekov2023}. The allowable defect types are a consequence of the system symmetry. For instance, consider a two-dimensional nematic liquid crystal where the averaged local orientation of the rod-like molecules is described by the director $\mathbf{n}$ and the angle $\varphi$, respectively. The topological charge of a defect $q$ can be determined by enclosing the defect in a loop $\gamma$ and calculating the change in director orientation around the loop \cite{KlemanLavrentovich2007}:
\begin{equation}
    q = \frac{1}{2\pi} \oint_\gamma \nabla \varphi \ dl.
\end{equation}
Topological defects in nematics have a minimum charge magnitude of $|q| = \sfrac{1}{2}$, reflecting the head-tail symmetry of their rod-like components. Nematic defects are disclinations, localized regions that violate orientational order \cite{KlemanLavrentovich2007,Alexander2012}. More complex, smectic liquid crystals also break translational symmetry by forming lamellae, introducing a density modulation on top of orientational ordering. Smectics can thereby support both disclinations and dislocations -- defects in the density periodicity \cite{KlemanLavrentovich2007,Kamien2006}. The ability to classify and control defects in more complex liquid crystal phases is necessary to apply them in nano- and bio-technologies \cite{Vroege2006,Querner2008,Zanella2011,Diroll2015,Hosseini2020,Hussain2021,Jehle2021}. 

Interest in smectics has been recently renewed due to advances in experimental techniques \cite{LopezLeon2011,Liang2011,Jeong2012,Kuijk2012,Serra2015,Coursault2016,Gim2017,Cortes2017,Repula2018,Blanc2023}, topological classification \cite{Chen2009,Kamien2016,Aharoni2017,Monderkamp2021,Hocking2022,Monderkamp2023,Severino2023}, and continuum modeling \cite{Pevnyi2014,Xia2021,Paget2023,Wensink2023}. Defects in molecular smectics have been stabilized with topological and geometrical constraints, but analyses have predominantly relied on mesoscopic modeling of the elastic energy $F$ \cite{LopezLeon2011,Liang2011,Jeong2012,Serra2015,Coursault2016,Gim2017}, given by:
\begin{equation}
    F = \int dV \left\{ \frac{B}{2} \left[ \frac{\partial u}{\partial z} - \frac{1}{2} \left( \frac{\partial u}{\partial x}\right)^2 \right]^2 + \frac{K}{2} \left( \frac{\partial^2 u}{\partial x^2} \right)^2\right\},
    \label{Eq:SmElasticity}
\end{equation}
where $V$ is the volume, $u(x,z)$ is the displacement of the smectic lamellae, and $B$ and $K$ are the layer compression and the layer bending moduli, respectively \cite{deGennes}. In the colloidal domain, theoretical efforts focused primarily on microscopic, classical density functional theory and Monte Carlo (MC) simulations to map out smectic order in the phase diagrams of hard rods \cite{Frenkel1988,Poniewierski1988,vanRoij1995,Graf1999,Savenko2004}. The phase diagrams have been probed experimentally with suspensions of rod-like viruses \cite{Wen1989,Grelet2014} and silica particles \cite{Kuijk2012,Kuijk2014-2}. Only this year has microscopic theory been connected to smectic elasticity (Eq.~\ref{Eq:SmElasticity}) \cite{Wensink2023}. Likewise, researchers have only begun to address smectic defects from a microscopic viewpoint. Using simulations and experiments, colloidal smectics have been confined within rectangles \cite{Cortes2017,Heras2006,Geigenfeind2015}, hexagons \cite{Monderkamp2021}, circles \cite{Heras2014,Pinto2017,Wittmann2021}, and annuli \cite{Wittmann2021,Armas2020,Wittmann2023}, establishing a correlation between the confining geometry and the defects formed. The defects were found to obey global, topological constraints \cite{Wittmann2021,Monderkamp2021,Wittmann2023,Monderkamp2023}. Yet, the exact role of local curvature in determining defect types remains an open question. 

\begin{figure*}
\includegraphics[width=0.85\textwidth]{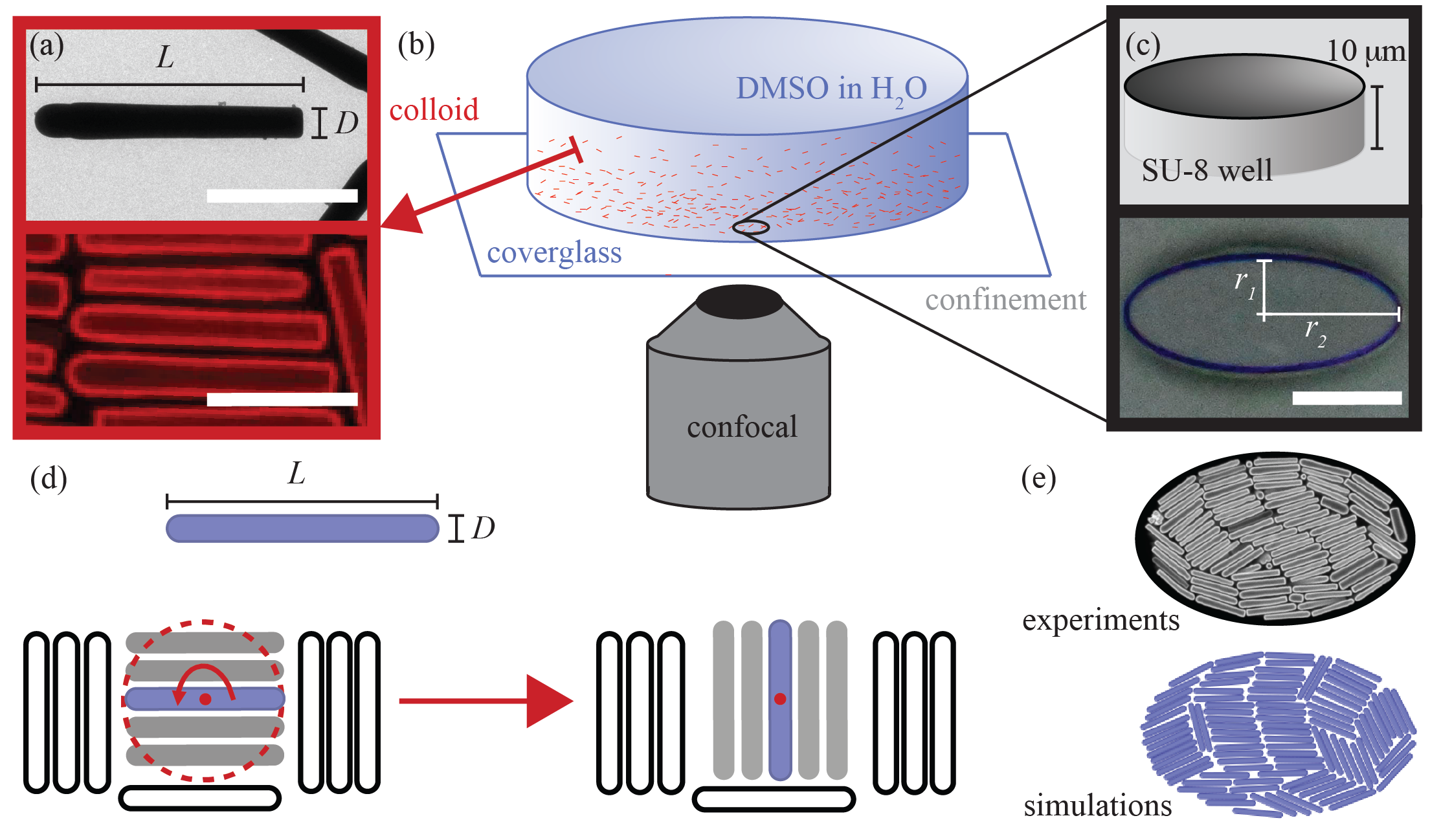}
\caption{(a) Transmission electron (top) and fluorescence confocal (bottom) micrographs of synthesized colloidal silica rods, with length $L=8.6\pm0.1$ {\textmu}m and diameter $D=1.16{\pm}0.02$ {\textmu}m. Scale bars are 5 {\textmu}m. (b) Schematic of experimental setup depicting the glass container filled with the silica rods suspended in 91 wt.-\% DMSO-in-water. The confinement wells are on the bottom coverglass, and the rods are sedimented into the wells. (c) Schematic and micrograph of an elliptical confinement with $r_1/r_2{=}0.4$, $r_1/L{=}3.0$, made of SU-8 photoresist with a height of 10 {\textmu}m. Scale bar is 50 {\textmu}m. (d) Representation of MC simulations with two-dimensional, discorectangles with end-to-end, length-to-diameter ratio $L/D=8$. To speed up simulations, we perform MC moves of square groups of rods, in addition to that of single rods. (e) Example experiment and simulation snapshots.}
\label{fig:AAF-ExpSetup}
\end{figure*}

For hard colloidal rods, the local curvature of a confining wall affects the liquid crystal \textit{anchoring}, which sets the rod alignment at a boundary \cite{Jerome1991}. Anchoring is a key tool in patterning molecular liquid crystals that remains underutilized for colloidal systems. For liquid crystal molecules, the anchoring can be selected by choosing the appropriate surface chemistry, but these methods are not applicable to larger-scaled systems. Since particle-based models rely predominantly on entropic interactions, only the geometry of the confining boundary can be tuned. The boundary geometry can alter excluded volume interactions to select an anchoring type \cite{Allen1999,Rodriguez-Ponce1999,Dijkstra2001,Barmes2004,Heras2005,Basurto2020}: either tangent to the wall (planar anchoring) or perpendicular to it (homeotropic anchoring). Planar anchoring at surfaces is commonly observed \cite{Cosentino-Lagomarsino-2003,Lewis2014,Garlea2016,Cortes2017,Wittmann2021}. On the other hand, homeotropic anchoring of hard rods has been examined in simulations and density functional theory frameworks but has yet to be explored experimentally \cite{Allen1999,Rodriguez-Ponce1999,Barmes2004,Heras2005,Cortes2017}. Achieving homeotropic anchoring is an open challenge that must be addressed to effectively design defects in colloidal liquid crystals. 

Here, with complementary experiments and simulations, we achieve anchoring and defect control of colloidal smectics by confining them within ellipses of varying sizes and eccentricity. The ellipse geometry creates a smooth variation of the local curvature, such that the critical curvature for anchoring transitions can be identified. Experimentally, we synthesize colloidal silica rods with fluorescently-labeled shells, which enable single-particle tracking of each rod using fluorescence confocal microscopy. We confine the silica rods by sedimentation into elliptical wells fabricated with photolithography. Using MC simulations, we model the system using a two-dimensional, monodisperse system of ``discorectangles". Working with an end-to-end, length-to-diameter ratio of $\sim8$, we study a smectic phase at packing fractions that exhibit meta-stable and short-range, quasi-tetratic ordering --- local regions with fourfold rotational symmetry. Note that the semicircles at the tips of the discorectangles are included in our definition of the rod length, which may deviate from other simulation studies. To analyze the structures, we examine nematic and tetratic order parameters, along with a network analysis of density maxima and minima. We identify features within the network that unambiguously distinguish disclinations, dislocations, and tetratic regions. In both simulations and experiments, we find that anchoring control from highly elliptical confinements stabilizes the smectic phase and creates dislocations. On the other hand, more circular confinements give rise to disclinations and tetratic ordering. Interestingly, differences in the number and types of defects between experiments and simulations point to changes in the smectic elasticity with the presence of polydispersity in the rod size. With this work, we demonstrate anchoring control of a colloidal smectic, broadening the tunability of defect states in larger-scaled, liquid crystal systems.

Our results are presented as follows: In Sec. II, we briefly describe our experimental and computational methodology. In Sec. III, we describe the nematic and tetratic order for our colloidal smectic, both in bulk and within elliptical confinements. We then determine how the rod anchoring varies with confinement curvature. In Sec. IV, we identify four main disclination states with varying confinement size and eccentricity. In the last section, Sec. V, we use minimum and maximum density networks in both experiments and simulations to analyze the number of disclinations and dislocations with varying ellipse geometry. We end by examining the effect of anchoring on metastable states.

\begin{figure*}
\includegraphics[width=0.9\textwidth]{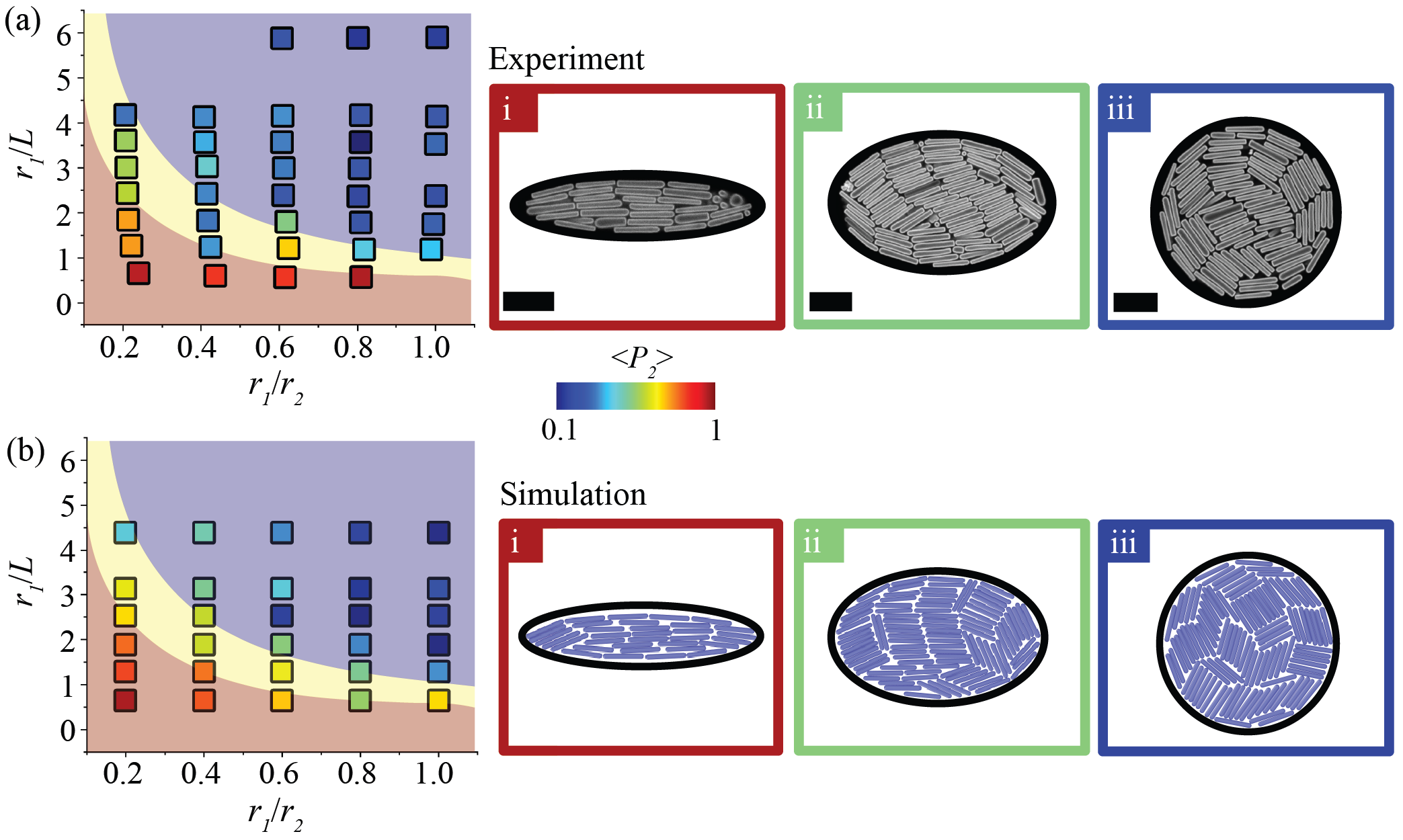}
\caption{Global nematic order parameter $\left< P_{2} \right>$ plotted for varying confinement shape $r_1/r_2$ and size $r_1/L$ for both (a) experiments and (b) simulations. The background colors denote different regimes, characterized by the smallest value of the radius of curvature $R_\text{min}$ for a given ellipse, where red is $R_\text{min}/L\leq0.5$, yellow is $0.5\leq R_\text{min}/L\leq1$, and blue is $R_\text{min}/L\geq1$. Experiment and simulation snapshots are shown on the right for confinement parameters of i) $r_1/r_2{=}0.2$, $r_1/L{=}0.6$, ii) $r_1/r_2{=}0.6$, $r_1/L{=}1.7$, and iii) $r_1/r_2{=}1.0$, $r_1/L{=}2.3$. The box color of each snapshot reflects the $\left< P_{2} \right>$ value. For experiments, the $\left< P_{2} \right>$ values are averaged over several snapshots and samples, and are approximately: i) $0.95$, ii) $0.52$, and iii) $0.20$. Scale bars are 10 {\textmu}m. Simulation snapshots have $\left< P_{2} \right>$ values of approximately i) $0.96$, ii) $0.53$, and iii) $0.16$.}
\label{fig:AAC-GlobalP2}
\end{figure*}

\section{\label{sec:Methodology}Methodology}

\subsection{Experimental methods}

In our experiments, we use fluorescently-labeled silica rods that are imaged using high-resolution, confocal laser scanning microscopy (Fig.~\ref{fig:AAF-ExpSetup}(a) and (b)). Details of the rod synthesis can be found in Appendix~\ref{App:ColloidalSynthesis}. Notably, each rod is labeled by a shell of fluorescent molecules, enabling single-particle tracking of each rod position and orientation, as well as particle-resolved characterization of smectic defects. To reduce the polydispersity of the obtained rods, the rod suspension was cleaned via centrifugation, details of which can be found in Appendix \ref{App:ColloidClean}. Our fluorescently-labeled silica rods have an average length of $L{=}8.6{\pm}0.1$ {\textmu}m and diameter of $D{=}1.16{\pm}0.02$ {\textmu}m, with polydispersities of $13{\%}$ and $20{\%}$, respectively (Fig.~\ref{fig:AAF-ExpSetup}(b)). Averaging over the length-to-diameter ratios of each measured rod gives $L/D{=}7.7{\pm}0.1$ (Fig.~\ref{fig:AAF-ExpSetup}(a)). The rod length-to-diameter ratio used in our study is smaller than those used in previous investigations \cite{Monderkamp2021, Cortes2017, Wittmann2021, Wittmann2023, Pinto2017, Armas2020} and gives rise to additional metastable structures beyond smectic layering, discussed further in Sec.~\ref{sec:Bulk ordering and anchoring}. 

To form a colloidal smectic, we concentrate the rods at the bottom of our samples via sedimentation. To match the refractive index of silica microparticles ($\sim1.45$), the rods are sedimented in 91 wt.-\% dimethylsulfoxide (DMSO, Sigma-Aldrich) in water, allowing for high-resolution confocal microscopy and single-particle analysis (Fig.~\ref{fig:AAF-ExpSetup}(b)). At the bottom of the samples are confining wells fabricated on ${\#}1$ cover glass (130-170 {\textmu}m thickness, VWR) using standard contact photolithography techniques, details of which can be found in Appendix \ref{App:Photolith}.

To alter the local curvature, we vary our elliptical confinements by changing their relative size and shape. The local curvature is set by the system size, measured by the ratio of the ellipse short-axis radius to the rod length ($r_1/L$), as well as the system shape, adjusted by $r_1/r_2$, the ratio of minor to major axes of the ellipse, respectively (Fig.~\ref{fig:AAF-ExpSetup}(c)). The ratio $r_1/r_2$ determines the eccentricity $e$ of a conic section, defined as $e^2 = 1 - (r_1/r_2)^2$, which, for an ellipse, takes on values between 0 (circular) and 1 (highly elliptical). The confinement size $r_1/L$ is varied from $0.6$ to $5.9$, in steps of $\sim0.6$. The confinement shape is varied with $r_1/r_2$ from $0.2$ to $1.0$, in steps of $0.2$. The global curvature of the confinement is kept fixed and is described by the Euler characteristic ($\chi$), a topological quantity that is invariant to smooth deformations. Following the Poincar{\'e}-Hopf theorem, the total topological charge $q$ is conserved, with $\sum q= \chi = +1$ \cite{Poincare,Hopf_theorem}. Further details on the topological charge conservation can be found in Sec.~\ref{sec:NetworkAnalysis}, where the network analysis is presented.

\subsection{\label{Simulation}Computational methods}

In our computational approach, we consider 2D monodisperse systems of rods modeled as hard spherocylinders of end-to-end, length-to-diameter ratio $L/D{=}8$ (Fig.~\ref{fig:AAF-ExpSetup}(d), top). As the particles are confined to lie in a plane, they can be more precisely described as ``discorectangles'', defined as a rectangle of length $L-D$ and width $D$ capped at each end by a semicircle of diameter $D$. The bulk phase behaviour is studied using constant pressure ($NPT$) MC simulations under periodic boundary conditions. The equation of state is obtained from compression of an isotropic phase and expansion of a perfect smectic-like phase, using $N{=}1000$ and $N{=}996$ particles, respectively.

In order to mimic the experimental systems, we perform MC simulations of particles confined within various two-dimensional elliptical boundaries. Particles are forced to reside within the elliptical regions by assuming a purely hard interaction with the external walls, thereby favoring planar surface anchoring. As we are interested in smectic-like structures, we fix the area fraction of the confined particles at ${\phi}{=}NA_{p}/A{=}0.81$, with $A_{p}{=}(L-D)D{+}{\pi}D^2/4$, the area of a hard particle and $A{=}{\pi}r_1r_2$, the area of the confining elliptical geometry. To obtain the desired value of ${\phi}{=}0.81$, where the smectic phase is stable in bulk, we slowly compress a low-density configuration with ${\phi}{<<}1$, initially placed in a scaled confining geometry.

To speed up the equilibration of the MC simulations, we implement, in addition to simple particle translation and rotation moves, MC moves in which square clusters of particles of dimensions approximately equal to the length of a single particle, are rotated by ${\pm}90^{\circ}$ (Fig.~\ref{fig:AAF-ExpSetup}(d), bottom). All simulations are run for at least $2.5{\times}10^{7}$ cycles. Unless stated otherwise, all the average quantities are measured over 1000 configurations selected from the last $10^6$ MC cycles. We then compare computational and experimental results for each system size and eccentricity (Fig.~\ref{fig:AAF-ExpSetup}(e)). 

\section{\label{sec:Bulk ordering and anchoring}Confinement and anchoring}

Using a smectic that exhibits metastable structures, we examine the influence of confinement on smectic ordering. To this end, we use rods with an aspect ratio of $L/D=8$ in simulations and $L/D=7.7{\pm}0.1$ in experiments. Simulations by Bates and Frenkel predicted that two-dimensional rods with an end-to-end, length-to-diameter ratio of $L/D\sim 8$ form a smectic phase at packing fractions of $\phi{\geq}0.75$ \cite{Frenkel2000}. Yet, there are subtleties in the pathways taken to form the smectic phase in simulations. The isotropic-to-smectic transition is not preceded by a uniaxial nematic state. In bulk, this effectively hinders nucleation of a smectic phase by direct compression from an isotropic phase~\cite{Frenkel2000,Ni2010}. Instead, (short-ranged) metastable tetratic ordering is commonly observed \cite{Frenkel2000,Ni2010,Pinto2017}. Bates and Frenkel find that even in simulations incorporating stack-rotation moves, a single-domain smectic phase can only be reached with \textit{expansion} to $\phi=0.75$ from a well-ordered and fully aligned crystal. Using rods with low aspect ratios $L/D\sim8$, we also expect to see metastable, tetratic ordering in our experiments and simulations.

To classify the nematic and tetratic order in our system, we use the 2D local orientational order parameter $P_{m}(\boldsymbol{r}){=}|\left<\exp(i m \theta_{j}) \right>_{\boldsymbol{r}}|$, with $\theta_{j}$ the angle formed by the orientation vector of the $j$-th particle ($\hat{\boldsymbol{u}}_{j}{=}(\cos \theta_{j}, \sin \theta_{j})^{T}$) with respect to an arbitrary fixed axis, the angular brackets $\left< \cdots \right>_{\boldsymbol{{r}}}$ denote an average over all particles which intersect a local circle around $\boldsymbol{r}$ with radius $r_c{=}4D$, and $m$ is an integer \cite{Monderkamp2021}. For $m{=}2$, $P_{m}(\boldsymbol{r})$ measures the degree of local nematic order, while for $m{=}4$ it gives the degree of local tetratic order. Both $P_2(\boldsymbol{r})$ and $P_4(\boldsymbol{r})$ take values ${\in}[0,1]$, where 0 and 1 indicate low and high orientational order around $\boldsymbol{r}$, respectively. We additionally calculate the global nematic and tetratic order parameter of the system $\left< P_{m} \right>$ by determining the order parameter over the entire system area. We use global order parameter values to characterize the degree of order in the system, while local order parameter values are used to identify disclinations. 

In Appendix~\ref{App:Bulk behavior, Bates and Frenkel}, we plot $\left<P_2\right>$ and $\left<P_4\right>$ with increasing packing fraction $\phi$ for our unconfined experimental and computational system. In simulations, the smectic phase is not accessible under compression. Instead, the system stabilizes high-density states characterized by perpendicularly aligned solid domains or states with local tetratic ordering, as expected from the work of Bates and Frenkel \cite{Frenkel2000} (low $\left<P_2\right>$ and moderate $\left<P_4\right>$ values at $\phi\sim0.75$, see Appendix~\ref{App:Bulk behavior, Bates and Frenkel}). Similarly, smectic ordering is only accessible with expansions from a perfectly aligned crystal phase to $\phi \geq 0.75$ (Fig.~\ref{fig:AAB-UnconfinedData}(c-iv)). Curiously, for unconfined silica rods under a sedimentation-diffusion equilibrium, we find tetratic ordering at $\phi\approx0.8$ that relaxes over time to predominantly smectic ordering with a few local, tetratic regions (Fig.~\ref{fig:AAB-UnconfinedData}(d-iii) and (d-iv)). We attribute the smectic ordering in our experiments to flow alignment and out-of-plane rod motion serving as ``local expansions" (see Appendix~\ref{App:Bulk behavior, Bates and Frenkel}). Over the several weeks where we observe \textit{confined} silica rods, we note that flow effects are minimal, but out-of-plane rod motion is still observed. 

At $\phi\approx0.75$ for both experiments and simulations, confinement appears to influence the bulk ordering, depending upon the confinement shape and size. In Fig.~\ref{fig:AAC-GlobalP2}, $\left<P_2\right>$ is plotted as a function of the system shape $r_1/r_2$ and size $r_1/L$. A reduction in $\left<P_2\right>$ is observed when moving from the bottom left region, low value of $r_1/r_2$ and small system size, to the top right, high value of $r_1/r_2$ and large system size. Generally, decreasing $r_1/r_2$, i.e. increasing confinement eccentricity, leads to greater $\left<P_2\right>$ values, indicating orientational order from the formation of well-aligned, smectic layers (Fig.~\ref{fig:AAC-GlobalP2}(a-i) and (b-i)). 

\begin{figure}[t]
\includegraphics[width=0.45\textwidth]{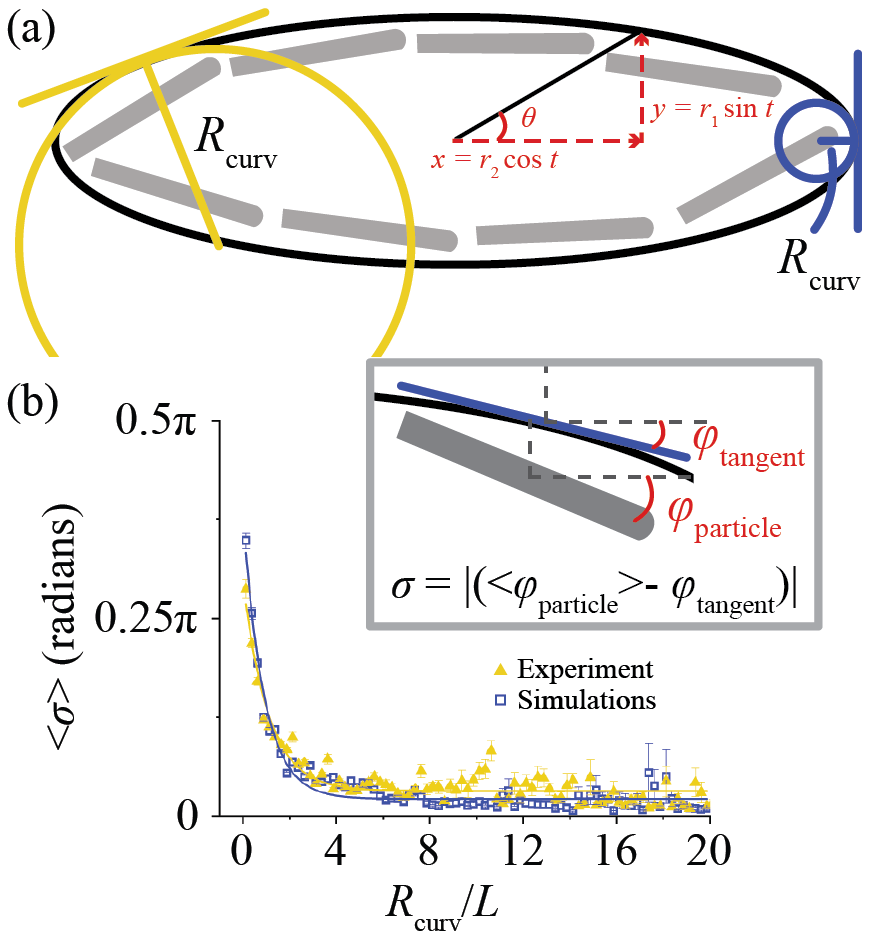}
\caption{(a) The radius of curvature $R_\text{curv}$ varies around an ellipse perimeter. Example radii of curvature at two different points on the ellipse boundary are drawn here with the associated tangent line at each point. The radius of curvature is given by Eq.~\ref{eq:EllipseRcurv}. The Cartesian axes are labeled in red, with the eccentric angle $\theta$ given by $\theta = \tan^{-1}(r_1/r_2 \tan t)$, where $t$ is the polar angle. (b) The average deviation $\left< \sigma \right>$ of the average particle orientation $\left< \varphi_\text{particle} \right>$ from the boundary tangent angle $\varphi_\text{tangent}$ (inset), is plotted as a function of $R_\text{curv}/L$. A transition from homeotropic (perpendicular) to planar (tangential) anchoring is observed with increasing $R_\text{curv}/L$. Both experimental (yellow triangles) and simulated data points (blue squares) are fitted to exponential decays.}
\label{fig:AAE-CurvatureAngleDevi}
\end{figure}

We attribute the increased smectic ordering with higher confinement eccentricity to the change in rod anchoring from planar to homeotropic near the ellipse vertices. An ellipse boundary is given by $x = r_2\cos t$, $y = r_1 \sin t$, where $x$ and $y$ are Cartesian coordinates, and $t$ is the polar angle, which varies from 0 to 2$\pi$. The radius of curvature, $R_\text{curv}$, for an ellipse varies around the perimeter, as illustrated in Fig.~\ref{fig:AAE-CurvatureAngleDevi}(a), and is given by: 

\begin{equation}
\label{eq:EllipseRcurv}
    R_\text{curv}(t){=}{\frac{{({r_1}^2\cos^2t+{r_2}^2\sin^2t})^{3/2}}{r_1 r_2}}.
\end{equation}

We define the smallest value of $R_\text{curv}$ for a given ellipse shape as $R_\text{min}$, located at the ellipse vertices and focus on how $\left< P_2 \right>$ varies with $R_\text{min}$. For $R_\text{min}/L{\leq}0.5$ (red region of graphs in Fig.~\ref{fig:AAC-GlobalP2}), we observe homeotropic anchoring at the ellipse ends, shown in Fig.~\ref{fig:AAC-GlobalP2}(a-i) and (b-i). The homeotropic anchoring gives rise to a single smectic domain, with a high $\left< P_2 \right>$ of ${\geq}0.7$. Moving towards intermediate $R_\text{min}$ values, ($0.5{\leq}R_\text{min}/L{\leq}1$, yellow region of graphs in Fig.~\ref{fig:AAC-GlobalP2}), we observe a reduction in $\left< P_2 \right>$ from $\sim0.7$ to $\sim0.5$. We attribute this reduction in $\left<P_2\right>$ to an increase in the proportion of rods tilting away from the boundary normal at the ellipse vertices, shown in Fig.~\ref{fig:AAC-GlobalP2}(a-ii) and (b-ii). Lastly, for confinements with $R_\text{min}/L{\geq}1$ (blue region of graphs in Fig.~\ref{fig:AAC-GlobalP2}), we find a relatively flat distribution of rod orientations, yielding $\left< P_2 \right>{\leq}0.4$. Examples are shown in Fig.~\ref{fig:AAC-GlobalP2}(a-iii) and (b-iii). We attribute the lower orientational order in the system to planar anchoring being obeyed around the entire confinement boundary. $R_\text{min}$ reduces below some critical value beyond which an anchoring transition from homeotropic to planar occurs.

To identify the critical radius of curvature $R^*_\text{curv}$ for an anchoring transition to arise, we analyze the average particle orientation at the boundary for all confinement geometries. The deviation of the average rod orientation angle $\left< \varphi_\text{particle} \right>$ from the boundary tangential angle $\varphi_\text{tangent}$ (see insert of Fig.~\ref{fig:AAE-CurvatureAngleDevi}(b)) is calculated around the ellipse perimeter for each confinement. At each point around the perimeter, we draw a circle with a radius twice the rod diameter $2D$ and average the rod orientation weighted by the area of each rod that is enclosed within the circle, to determine $\left< \varphi_\text{particle} \right>$. We also calculate $R_\text{curv}$ at each increment using Eq.~\ref{eq:EllipseRcurv}. The angle deviation $\sigma = |(\left< \varphi_\text{particle} \right> - \varphi_\text{tangent})|$ is binned and then averaged as a function of $R_\text{curv}/L$ across all confinement samples, plotted in Fig.~\ref{fig:AAE-CurvatureAngleDevi}(b). Additional details can be found in Appendix~\ref{App:RcurvCalc}. We fit the average angle deviation $\left<\sigma\right>$ as a function of $R_\text{curv}/L$ to an exponential decay, finding $\left< \sigma \right> \approx (0.27) \exp{[-0.93(R_\text{curv}/L)] + 0.03}$ for experiments and $\left< \sigma \right> \approx (0.38) \exp{[-1.18(R_\text{curv}/L)] + 0.02}$ for simulations. Defining planar anchoring as a rod orientation $\pm 15$ degrees from the tangent ($\left< \sigma \right> = 0.08\pi$), we find the critical radius of curvature $R^{*}_\text{curv}/L \approx 1.9$ for experiments and $R^{*}_\text{curv}/L \approx 1.6$ for simulations. 

$R^{*}_\text{curv}$ being approximately the rod length $L$ suggests that the excluded area of the rods changes with boundary curvature. With varying curvature, the rods switch their orientation with respect to the boundary to pack more efficiently. Instead of packing end-to-end as with planar anchoring, more rods can fill the area near the boundary when the rods stack side-to-side, yielding homeotropic anchoring. The trade-off in minimizing the rod-boundary excluded area with the excluded area with neighboring rods appears to occur between $L$ and $2L$. Our findings show that boundary curvature determines the anchoring condition for hard rods. 

\section{\label{sec:SmStructure}Smectic disclination structure}

\begin{figure*}
\includegraphics[width=0.85\textwidth]{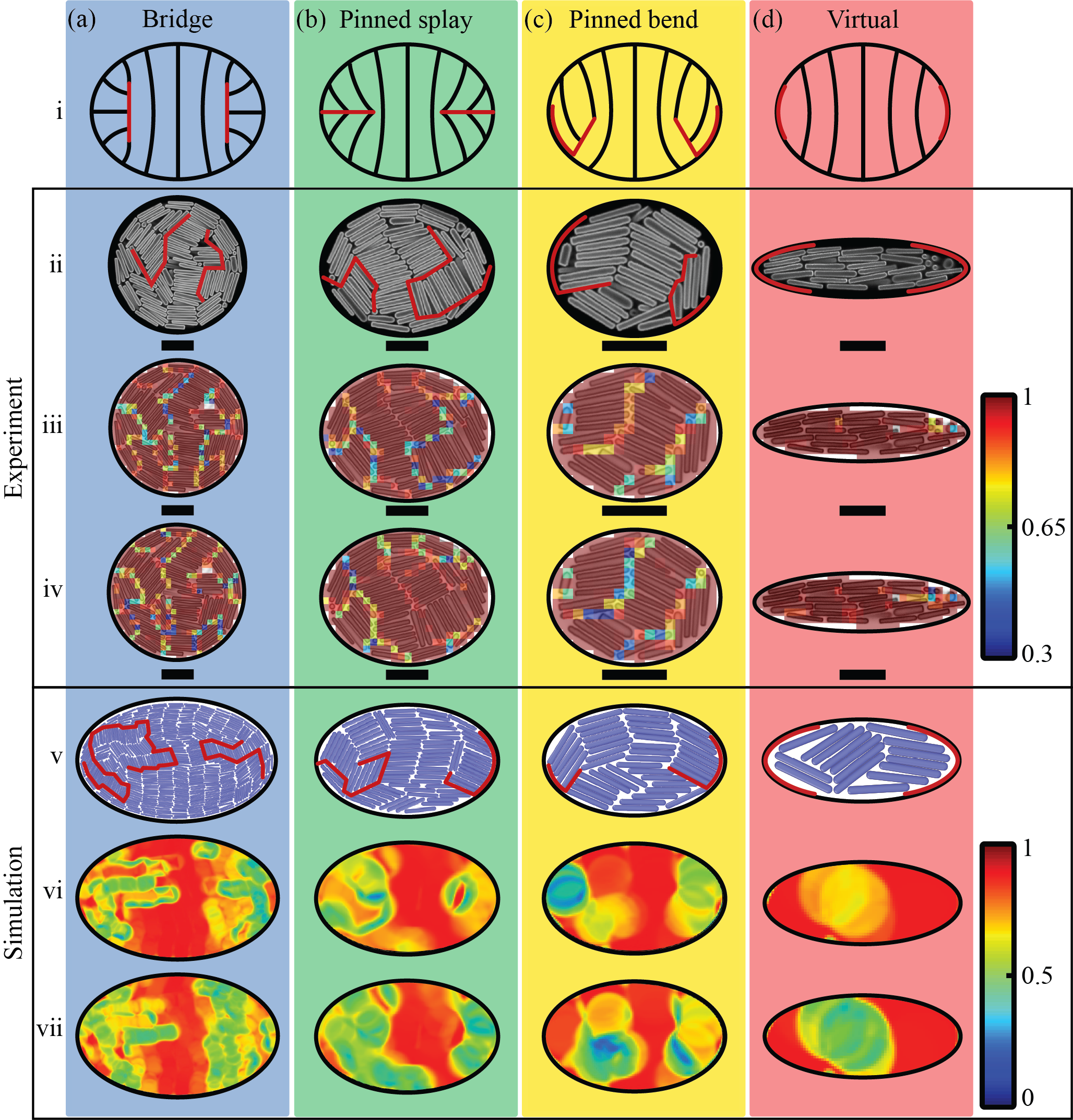}
\caption{(a) Bridge ($\mathcal{B}$), (b) pinned splay ($\mathcal{PS}$), (c) pinned bend ($\mathcal{PB}$), and (d) virtual ($\mathcal{V}$) disclination states observed for varying confinements. i) Idealized structures are shown with black lines representing smectic layers and red lines representing disclinations. Experiment ii) snapshots, iii) local $P_2$ mapping, and iv) local $P_4$ mapping are shown (scale bars are 10 {\textmu}m). Simulation v) snapshots, vi) local $P_2$ mapping, and vii) local $P_4$ mapping are also shown.}
\label{fig:FigAAA-SMStructureStates}
\end{figure*}

With confinement eccentricity setting the hard rod anchoring, we examine the effect of anchoring on the smectic disclination structure. The defects in the system are influenced by the confinement's local curvature (the size and eccentricity) as well as its global curvature (the fixed Euler characteristic $\chi = +1$). We first identify disclinations in experiments and simulations using local mapping of the nematic order parameter $P_2$ (Fig.~\ref{fig:FigAAA-SMStructureStates}). Our MC simulations are run for at least $2.5\times10^7$ cycles, while our experiments are observed periodically over a period between 1 and 5 months. In analyzing the smectic layer conformation, we ignore interstitials - single rods that orient parallel to the layers and are located in between them \cite{vanRoij1995}. Across the varying confinement sizes and shapes, the observed disclination configurations can be categorized by four states that we identify as the bridge ($\mathcal{B}$), pinned splay ($\mathcal{PS}$), pinned bend ($\mathcal{PB}$), and virtual ($\mathcal{V}$) states.

Starting at large confinements with low eccentricity ($r_1/L{\geq}2.4$, $r_1/r_2{\geq}0.6$), the $\mathcal{B}$ state is observed and is characterized by two antipodal disclinations, disconnected from the boundary, typically with a large central smectic domain (Fig.~\ref{fig:FigAAA-SMStructureStates}(a)) \cite{Monderkamp2021, Wittmann2021}. Planar anchoring is obeyed around the entire confinement boundary. 

For intermediate confinement sizes ($1.2{ \leq}r_1/L{\leq}2.4$), two distinct states are observed: the $\mathcal{PS}$ and $\mathcal{PB}$ states. Both are similar to the $\mathcal{B}$ state, having two, antipodal disclination lines and a central smectic domain. The distinction between these states lies in the adsorption of the disclination line to the boundary. The degree of disclination adsorption varies with the rod anchoring. The $\mathcal{PS}$ state has disclination lines adsorbed at approximately a single point on the boundary (Fig.~\ref{fig:FigAAA-SMStructureStates}(b)) and has predominantly planar anchoring, while the $\mathcal{PB}$ state has linear portions of the disclinations adsorbed (Fig.~\ref{fig:FigAAA-SMStructureStates}(c)) with larger regions of homeotropic anchoring. The $\mathcal{PS}$ state requires the director, which describes the rod orientation, to splay at the point of disclination adsorption. In the $\mathcal{PB}$ state, the smectic lamellae follow the curvature of the boundary and are bent without director distortion.

We attribute the rarity of the $\mathcal{PS}$ state to differences in elastic energy contributions of director splay and layer bend for the elastic constant $K$ in Eq.~\ref{Eq:SmElasticity} for hard-rod smectics. Wensink and Grelet explored these differences in the elastic response of hard-rod smectics with a recent work that connects microscopic theory to mesoscale elasticity \cite{Wensink2023}. By extending the work of Straley \cite{Straley1976} to smectics,  they find that the elastic modulus for smectic layer bending is typically two orders of magnitude lower than that of splay in the director for hard-rod smectics. The elastic constant $K$ in Eq.~\ref{Eq:SmElasticity} is therefore dominated by director splay deformations. In other words, for colloidal smectics, layer bending without director splay has a minimal elastic energy penalty, making the $\mathcal{PB}$ state more energetically preferred than the $\mathcal{PS}$ state.

Lastly, for extreme confinements ($r_1/L{\leq}1.2$), the $\mathcal{V}$ state is observed (Fig.~\ref{fig:FigAAA-SMStructureStates}(d)), characterized by the absence of disclinations in the bulk. In certain confining geometries of intermediate size and eccentricity, composite states $\mathcal{C}$ are observed that combine two of the above mentioned states. We denoted which states are part of the composite using super- and sub-scripts: for example, a composite structure of $\mathcal{B}$ and $\mathcal{PS}$ is denoted $\mathcal{C}_{\mathcal{B}}^{\mathcal{PS}}$.

\begin{figure}
\includegraphics[width=0.48\textwidth]{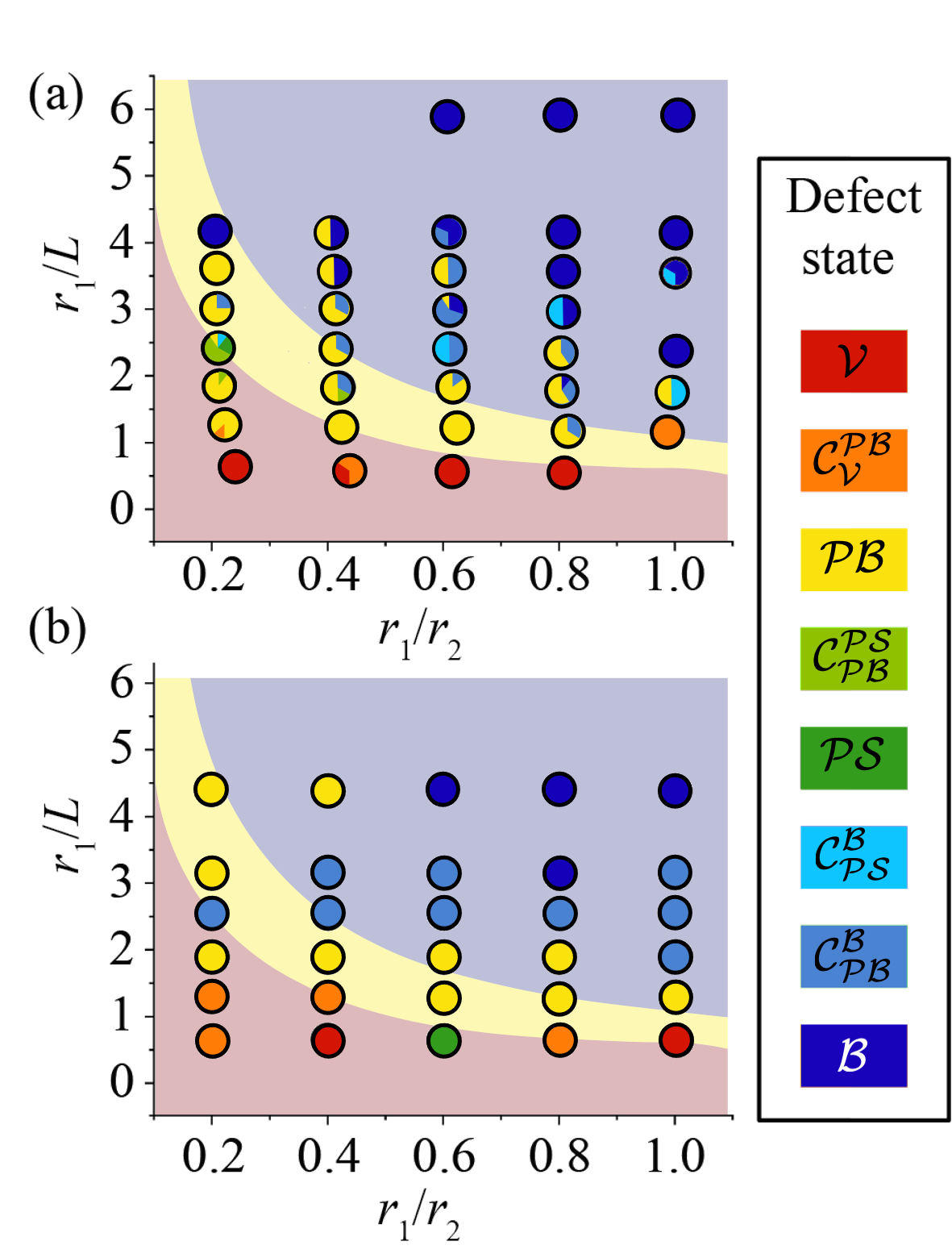}%Single column figure
\caption{Disclination states are plotted for varying confinement size ($r_1/L$) and shape ($r_1/r_2$), for both (a) experiments and (b) simulations. A transition from the $\mathcal{V}$ to the $\mathcal{B}$ state is observed with increasing confinement size and decreasing eccentricity (i.e. increasing $r_1/r_2$), passing through intermediate $\mathcal{PB}$, $\mathcal{PS}$, and composite states. The background colors again denote different regimes, characterized by the smallest value of the radii of curvature $R_\text{min}/L$ for a given ellipse, where red is $R_\text{min}/L\leq0.5$, yellow is $0.5\leq R_\text{min}/L\leq1$, and blue is $R_\text{min}/L\geq1$. The $R_\text{min}/L$ regimes demarcate the shifts in disclination states, indicating a relationship between the curvature-imposed anchoring and the disclination state.}
\label{fig:FigAAD-StrcutrePhaseDiagram}
\end{figure}

The disclination structure depends on the confinement geometry ($r_1/r_2$ and $r_1/L$) and thus the imposed anchoring. To visualize this dependence, we plot the disclination state with varying $r_1/r_2$ and $r_1/L$ in Fig.~\ref{fig:FigAAD-StrcutrePhaseDiagram}. In certain cases, the disclination state varies between samples or relaxes over time, so for these instances, the ratio of observed structures is plotted.

At high eccentricity and confinement, we observe the $\mathcal{V}$ state (bottom left of Fig.~\ref{fig:FigAAD-StrcutrePhaseDiagram}(a) and (b)). With larger confinements and decreasing eccentricity, the system configuration moves through the $\mathcal{PB}$/$\mathcal{PS}$ states and ends with the $\mathcal{B}$ state at the largest and most circular confinements ((top right of Fig.~\ref{fig:FigAAD-StrcutrePhaseDiagram}(a) and (b)). Similar to plots in Fig.~\ref{fig:AAC-GlobalP2}, $R_\text{min}$ becomes greater with increasing $r_1/L$ and $r_1/r_2$ and is colored from red ($R_\text{min}/L \leq 0.5$) to yellow ($0.5 \leq R_\text{min}/L \leq 1$) to blue ($R_\text{min}/L \geq 1$). The increase in $R_\text{min}$ alters the anchoring condition from homeotropic to planar once $R^*_\text{curv}/L\approx 1$ is reached. The anchoring change directly affects the disclination states. The homeotropic-to-planar anchoring transition that arises with reduced boundary curvature produces disclinations ($\mathcal{V}$ state transitions to $\mathcal{PB}$/$\mathcal{PS}$ states across $R_\text{min}/L=0.5$ from red to yellow regions) and de-pins them from the boundary at $R_\text{min}/L\geq R^*_\text{curv}/L=1$ (blue regions), pushing the disclinations into the bulk ($\mathcal{B}$ state).

The disclination state of a colloidal smectic relates directly to the dependence of anchoring on confinement curvature. Disclination de-pinning was similarly observed in the hard-rod system of Monderkamp \textit{et al.} when their confinement had rounded corners \cite{Monderkamp2021}. Here, we make explicit the connection between curvature, anchoring, and disclination pinning to a surface. The anchoring condition of a colloidal liquid crystal, controlled by the confinement curvature, determines the disclination state. 

\section{\label{sec:NetworkAnalysis}Network analysis}

We have thus far established the connection between confinement geometry, anchoring, and disclinations --- defects in the orientational order identified by low values in the nematic order parameter ($P_2$). Yet, smectics also have positional order and thereby have dislocation defects. Dislocations require an identification method that reflects the broken translational symmetry of the smectic phase. Therefore, in order to analyze the effect of confinement geometry on dislocations, we leverage the single-rod resolution of our system to draw networks of density maxima and density minima using a method recently introduced by Monderkamp \textit{et al.} \cite{Monderkamp2023}. 

\begin{figure}
\includegraphics[width=0.45\textwidth]{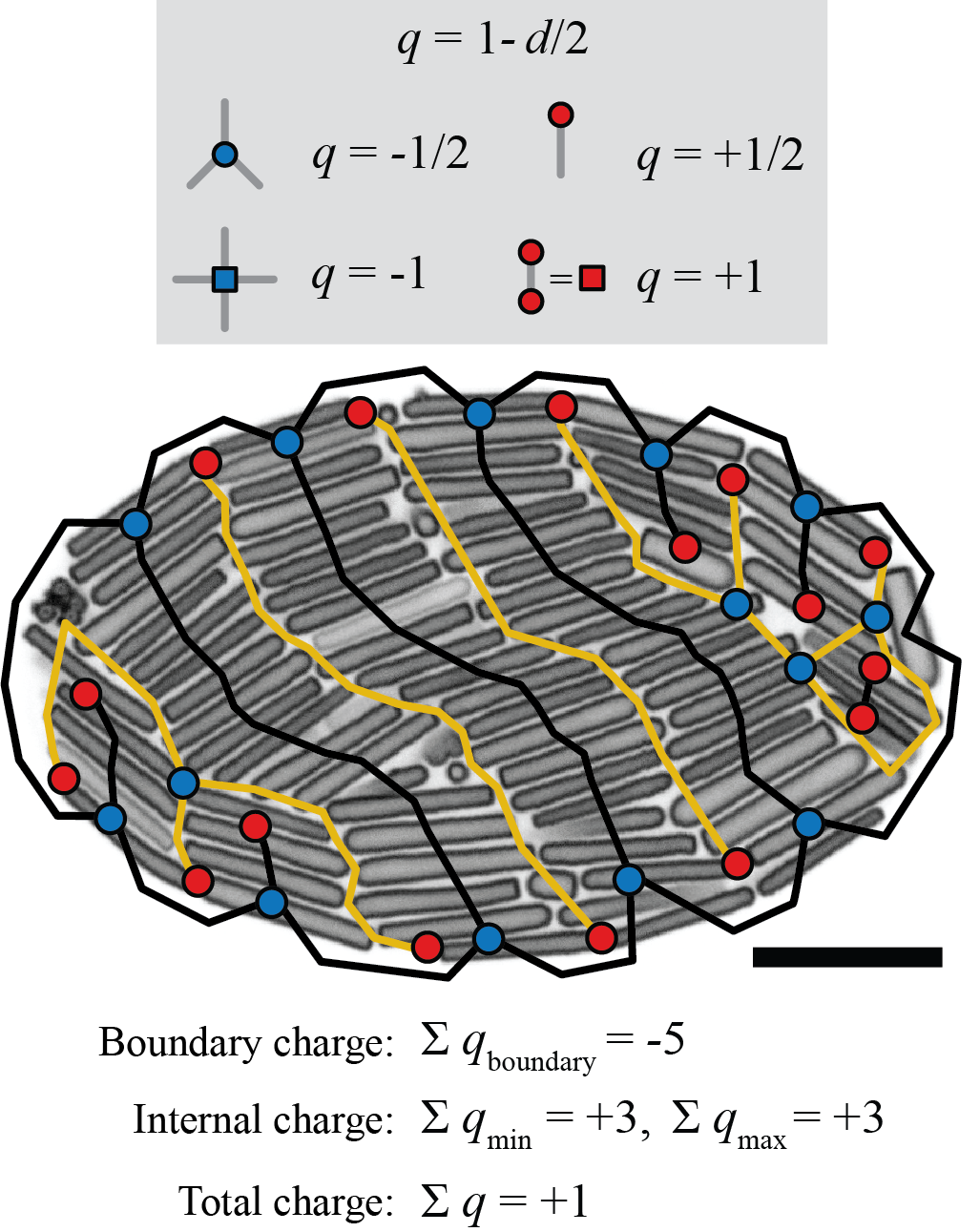}%Single column figure
\caption{Network analysis for confinement with $r_1/r_2{=}0.6$, $r_1/L{=}1.8$. Maximum density (black lines) and minimum density (yellow lines) networks are drawn, with positively charged vertices marked by red dots and negatively charged vertices marked by blue dots. The topological charge, $q$, for each vertex is given by $q = 1-d/2$, where $d$ is the number of lines connecting to the vertex. The boundary charges ($\sum q_\text{boundary}$) with the internal minimum ($\sum q_\text{min}$) and maximum ($\sum q_\text{max}$) density network charges sum to 1 and equals the Euler characteristic for an ellipse, as expected from the Poincar\'{e}-Hopf theorem. Scale bar is 10 {\textmu}m.}
\label{fig:FigAAI-NetworkAnalysisExample}
\end{figure}

Briefly, each network consists of a set of vertex points identifying rod centers or rod ends, from which a network is generated via a Delaunay triangulation of all points. The boundary can be arbitrarily assigned to either the minimum density (rod ends) or maximum density network (rod centers). Here, we choose to always assign the boundary to the maximum density network. Next, the minimum and maximum density networks are then separated and empty loops are collapsed to single vertex points or lines, resulting in two, interwoven, but separate minimum (yellow lines, Fig.~\ref{fig:FigAAI-NetworkAnalysisExample}) and maximum (black lines, Fig.~\ref{fig:FigAAI-NetworkAnalysisExample}) density networks. Similar to Sec.~\ref{sec:Bulk ordering and anchoring}, we exclude interstitials from our analysis. A detailed protocol for the network generation can be found in Appendix \ref{App:Network Generation Protocol}.

The maximum and minimum density networks carry topological charge. The total topological charge of a network $\sum q$ can be calculated by summing the charges of every vertex point on the network, where the charge $q$ is set by the number of adjacent connected points $d$: $q{=}1{-}d/2$. For elliptical confinement with the boundary assigned to either the minimum density or the maximum density network, the system follows the Poincar\'{e}-Hopf theorem with ${\sum}q{=}{\chi}{=}{+}1$. We illustrate the topological charge conservation by highlighting an example in Fig.~\ref{fig:FigAAI-NetworkAnalysisExample}. The boundary has ten, -1/2 point defects, giving a total boundary charge $\sum q_\text{boundary} = -5$. Both the internal maximum density and minimum density networks each have positively and negatively charged vertices that total to +3: $\sum q_\text{min} = +3$ and $\sum q_\text{max} = +3$. As expected, the total charge $\sum q = \sum q_\text{boundary} + \sum q_\text{min} + \sum q_\text{max} = +1$, which equals the Euler characteristic for an ellipse, $\chi = 1$. More discussion on the topological charge conservation of the minimum and maximum density networks of smectics can be found in the recent work of Monderkamp \textit{et al.} \cite{Monderkamp2023}.

However, beyond topological charge conservation, the networks also have features that uniquely distinguish topological defect types. In this section, we show how vertex charges and loops in the network can be used to quantify the amount of disclinations and dislocations, as well as to identify local tetratic ordering. We then use these network features to compare how the confinement geometry influences the formation of dislocations and metastable, tetratic domains.

\subsection{Network charges: disclinations and dislocations}

\begin{figure}[b]
\includegraphics[width=0.4\textwidth]{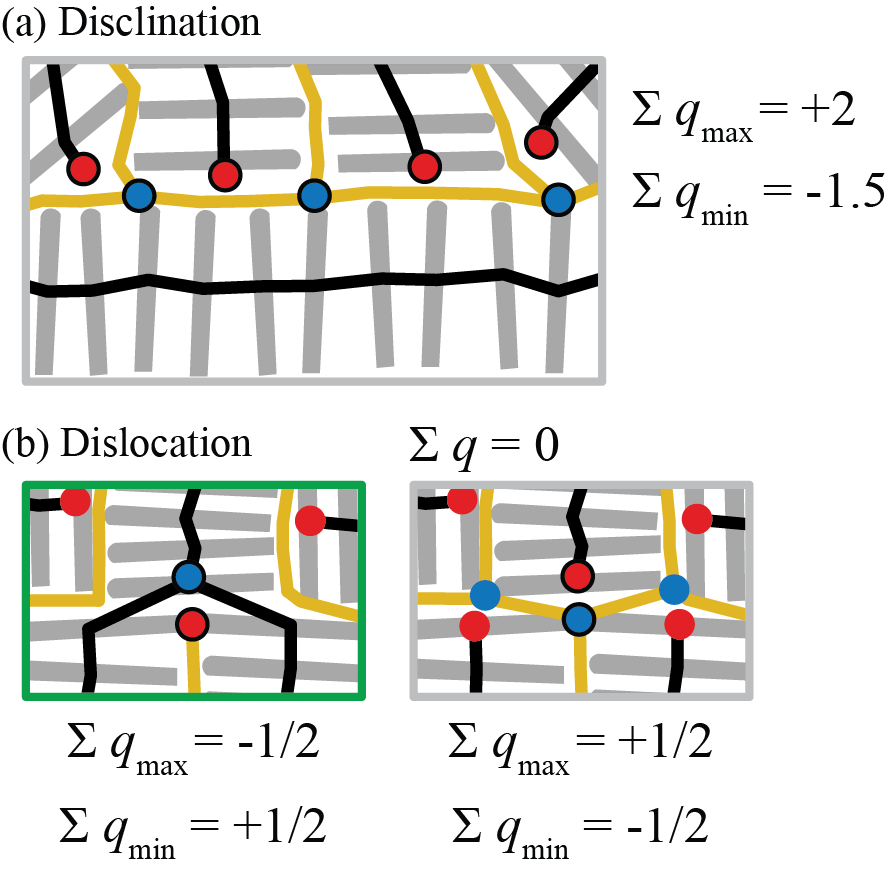}%Single column figure
\caption{Network topology associated with (a) disclinations and (b) dislocations. (a) Disclinations produce positive charge in the maximum density network and negative charge for the minimum density network and have total charge of $\sum q = +1/2$. (b) Dislocations have a degeneracy in the assignment of network charges. They can be represented by a split/merge in either the maximum density or minimum density networks. Each operation equivalently results in a total topological charge of zero. To distinguish dislocations from disclinations using network charges, we always assign the network split/merge to the maximum density network, highlighted by the green box (left).}
\label{fig:FigAAJ-NetworkAnalysisDefects}
\end{figure}

The vertex topological charges on the minimum density and maximum density networks can be used to locate disclinations and dislocations. In Fig.~\ref{fig:FigAAJ-NetworkAnalysisDefects}, we show schematics of a disclination (Fig.~\ref{fig:FigAAJ-NetworkAnalysisDefects}(a)) and a dislocation (Fig.~\ref{fig:FigAAJ-NetworkAnalysisDefects}(b)), with the maximum density networks labeled in black and the minimum density networks labeled in yellow. Positively charged vertices are represented by red dots, while negatively charged vertices are represented by blue dots. 

\begin{figure*}
\includegraphics[width=0.8\textwidth]{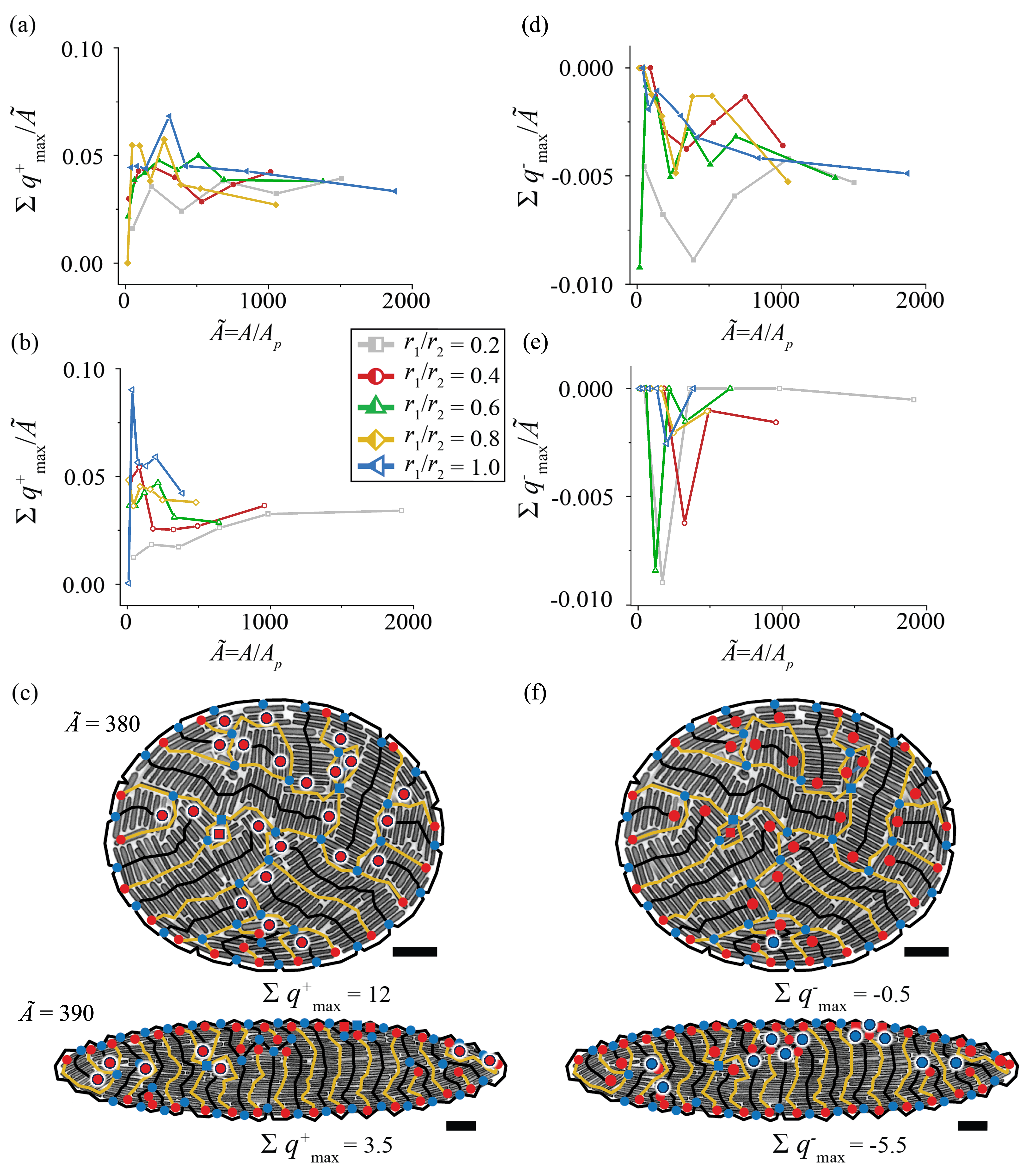}%Single column figure
\caption{Experiment (a, d) and simulation (b, e) results of the positive (a, b) and negative (d, e) network charge for the maximum density network $\sum q_\text{max}$, divided by the normalized confinement area $\widetilde{A} = A/A_p$, where $A$ denotes the confinement area and $A_p$ the particle area. $\sum q_\text{max}/\widetilde{A}$ is then plotted against $\widetilde{A}$. Positive maximum density network charge $\sum q^{+}_\text{max}$ is associated with disclinations, while negative maximum network charge $\sum q^{-}_\text{max}$ is associated with dislocations. Varying confinement shapes ($r_1/r_2$) are plotted in different colors, shown in the center legend. Filled data points are experiments, while open points are simulations. Experiment snapshots for $r_1/r_2{=}0.8$, $r_1/L{=}3.6$ (c and f, top) and $r_1/r_2{=}0.2$, $r_1/L{=}1.9$ (c and f, bottom) demonstrate that, for comparable $\widetilde{A}$, the confinement with higher eccentricity has fewer internal $q^{+}_\text{max}$ (red dots, highlighted by a white outline), indicating disclination suppression, as well as more internal $q^{-}_\text{max}$ (blue dots, highlighted by a white outline), indicating dislocation promotion. Scale bars are 10 {\textmu}m.}
\label{fig:FigAAG-MaxNetworkDefects}
\end{figure*}

In the vicinity of a disclination, the maximum density network gains positive charges, while the minimum density network gains negative charges. In regions with a discontinuous change in the rod orientation, the maximum density network must end at a single point, yielding a positive charge. Likewise, the minimum density network branches, forming a negative charge. The number of positive vertices on the maximum density network (or similarly, negative vertices on the minimum density network) can be used to quantify the amount of disclinations in a system. Disclinations have an excess topological charge of $+1/2$ in the network, as shown in Fig.~\ref{fig:FigAAJ-NetworkAnalysisDefects}(a). This is reflected in how the $\mathcal{B}$, $\mathcal{PS}$, and $\mathcal{PB}$ states are made up of two disclinations (Fig.~\ref{fig:FigAAA-SMStructureStates}), with a total system charge of +1, as expected from the Poincaré-Hopf theorem.

On the other hand, for dislocations, the network charges from both the minimum and maximum density networks total to zero, as illustrated in Fig.~\ref{fig:FigAAJ-NetworkAnalysisDefects}(b). A smectic dislocation is characterized by either a splitting of one layer into two, or likewise a merging of two layers into one, resulting in a discontinuity in the positional order. That a dislocation in smectics can be viewed as either a merger or a branching of layers is reflected in how the minimum and maximum density networks are drawn. We demonstrate the ambiguity in the network representation of dislocations in Fig.~\ref{fig:FigAAJ-NetworkAnalysisDefects}(b). The change in layer number from a disclination can be equally represented in the network as either branching in the maximum density network (Fig.~\ref{fig:FigAAJ-NetworkAnalysisDefects}(b), left) or equivalently, branching in the minimum density network (Fig.~\ref{fig:FigAAJ-NetworkAnalysisDefects}(b), right). To allow for a quantitative distinction between disclinations and dislocations using network charges, we choose to always represent a dislocation by branching the \textit{maximum} density network. With this convention for dislocations, negative vertices in the internal, maximum density network uniquely identify dislocations. The quantity of negative charge in the maximum density network then serves as a measure for the amount of dislocations. 

To determine the amount of disclinations and dislocations with varying confinement geometry, we plot the total internal positive and negative charges on the \textit{maximum} density network separately, $\sum q_\text{max}^{+/-}$ for $r_1/r_2$ varying from 0.2 to 1.0 against the system size in Fig.~\ref{fig:FigAAG-MaxNetworkDefects}. The system size is measured by the confinement area $A$ normalized by the area $A_p$ of a single rod: $\widetilde{A}=A/A_p$. The total positive charges $\sum q^{+}_\text{max}$ quantify the extent of disclinations in the system, while the total negative charges $\sum q^{-}_\text{max}$ quantify the number of dislocations. Experimental data is plotted in Fig.~\ref{fig:FigAAG-MaxNetworkDefects}(a) and (d), while simulated data is plotted in Fig.~\ref{fig:FigAAG-MaxNetworkDefects}(b) and (e). In general, the plots of Fig.~\ref{fig:FigAAG-MaxNetworkDefects} show $\sum q^{+}_\text{max}$ and $\sum q^{-}_\text{max}$ both increasing with system size and scaling with the area. Normalizing both of these quantities with $\widetilde{A}$ shows a generally flat distribution at larger system sizes, indicating that the length of disclinations and the number of dislocations grow with system size. 

We focus first on disclinations and $\sum q^{+}_\text{max}$ in Fig.~\ref{fig:FigAAG-MaxNetworkDefects}(a) and (b). Confinements with high eccentricity ($r_1/r_2{\leq}0.4$) produce fewer positive charges for a given area in both experiments and simulations. This result agrees with our findings in Fig.~\ref{fig:FigAAD-StrcutrePhaseDiagram}, where higher eccentricity with decreasing $r_1/r_2$ leads to boundary-adsorbed disclinations ($\mathcal{PS}$ and $\mathcal{PB}$ states) that disappear in the $\mathcal{V}$ state with smaller system sizes. Increasing the confinement eccentricity increases homeotropic anchoring near the ellipse vertices, which produces fewer/shorter disclinations in the bulk. Fig.~\ref{fig:FigAAG-MaxNetworkDefects}(c) exemplifies the decrease in positive internal vertices, marked by highlighted red dots, signifying a reduction in disclination amount with increased confinement eccentricity.

Turning towards dislocations and $\sum q^{-}_\text{max}$ in Fig.~\ref{fig:FigAAG-MaxNetworkDefects}(d) and (e), we now see major differences between experiments and simulations. Experiments show a larger amount of dislocations across the varying confinement geometries compared to simulations. We attribute this disparity in dislocation number to polydispersity in the rods used in experiments that is not present in simulations. In Appendix~\ref{app:Polydispersity}, we detail how modeling with polydisperse rods in simulations leads to larger magnitudes of $\sum q^{-}_\text{max}$ and more dislocations. Rods with lengths deviating from the average layer spacing appear to promote shifts in the smectic layering, resulting in dislocations. 

Despite the differences between experiments and simulations, the data shows that confinement geometry affects dislocation formation in confined smectics. Confinements with a higher eccentricity ($r_1/r_2{\leq}0.4$) have more negative internal charge, $\sum q^{-}_\text{max}$, showing an increase in dislocations. In experiments, confinements with $r_1/r_2=0.2$ have more dislocations compared to other confining geometries, and in simulations, $r_1/r_2=0.2$, 0.4, and 0.6 all have more dislocations than the more circular confinements. As shown by the examples in Fig.~\ref{fig:FigAAG-MaxNetworkDefects}(c) and (f), more elliptical confinements have a large smectic domain that can support dislocation formation, while more circular confinements have almost no dislocations and multiple domains separated by disclinations. 

The large smectic domain in confinements with high eccentricity supports the creation of dislocations due to boundary curvature. Consider a confinement ellipse with $r_1/r_2{=}0.2$ and $r_1/L{=}2.45$. Using the boundary charges from the network analysis (see Appendix~\ref{App:SmecticLayerSpacing}), we determine the average smectic layer spacing $\lambda$ to be $\lambda = 1.1L$. The number of possible layers $m$ that can fit across the center of the ellipse is $m_\text{center}{=}2{r_2/\lambda}=22.3$. On the other hand, the number of layers that can fit along the bottom of the confining boundary is $m_\text{bottom}{=}c/(2\lambda){=}23.4$. The incommensurate number of layers from the center of the system to the boundary leads to geometrical frustration that is relieved by the Helfrich-Hurault instability \cite{Blanc2023} and the nucleation of dislocations. The long-range smectic ordering, the incommensurate number of layers from the curved boundary, and the rod polydispersity all lead to the prevalence of dislocations in highly elliptical confinements.

Using topological charges from maximum density and minimum density networks, we find disclinations forming in circular confinements with planar anchoring, in line with our findings with $\left<P_2\right>$ (Sec.~\ref{sec:SmStructure}). Notably, the network charge analysis enables the quantification of dislocations. We find dislocations forming in confinements with high eccentricity, due to homeotropic anchoring in high curvature regions creating long-ranged smectic order.

\subsection{Network loops and tetratic ordering}
\begin{figure}[b]
\includegraphics[width=0.45\textwidth]{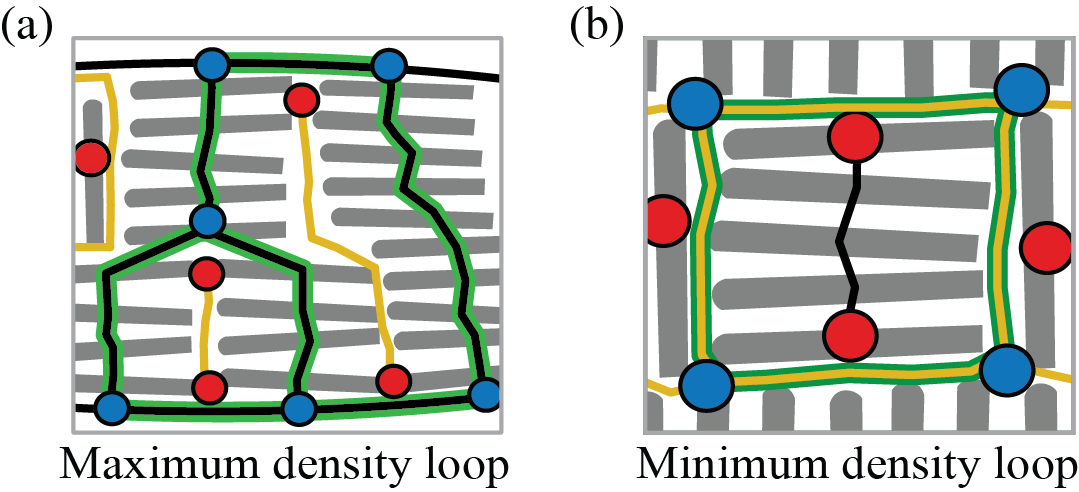}%Single column figure
\caption{Schematics showing (a) closed loops of the maximum density network corresponding to well-organized smectic-layers and (b) closed loops of the minimum density network corresponding to tetratic regions.}
\label{fig:FigAAK-MaxNetworkLoopsSchematic}
\end{figure}

\begin{figure*}
\includegraphics[width=0.8\textwidth]{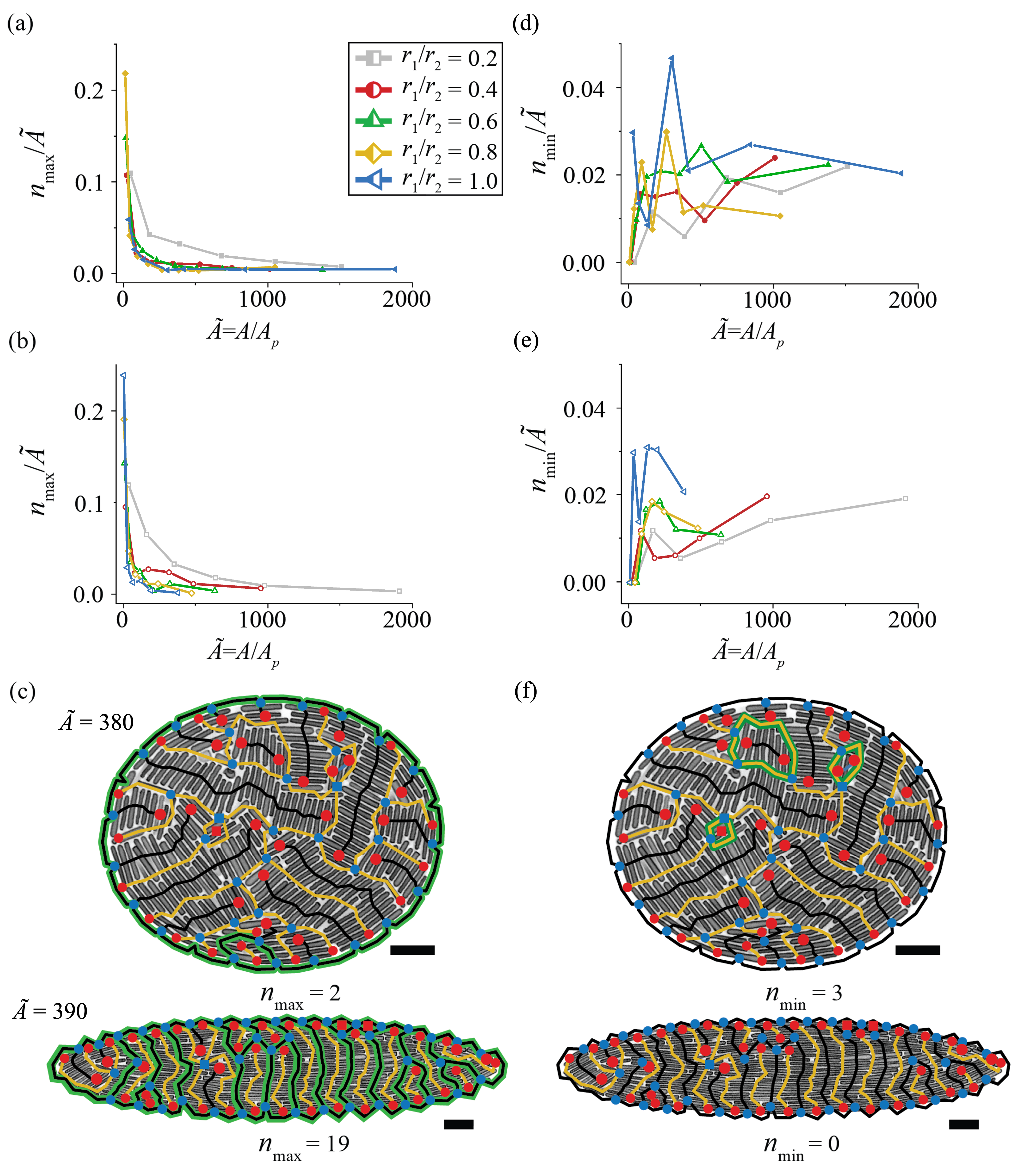}%Single column figure
\caption{Experiment (a, d) and simulation (b, e) results of the maximum density loops $n_\text{max}$ (a, b) and minimum density loops $n_\text{min}$ (d, e), divided by the normalized confinement area $A_p$: $\widetilde{A} = A/A_p$, where $A$ denotes the confinement area and $A_p$ the particle area. $n/\widetilde{A}$ is then plotted against $\widetilde{A}$. Maximum density network loops are associated with ordered smectic layers, while minimum density network loops are associated with tetratic regions. Varying confinement shapes ($r_1/r_2$) are plotted in different colors, shown in the top legend. Filled data points are experiments, while open points are simulations. Experiment snapshots for $r_1/r_2{=}0.8$, $r_1/L{=}3.6$ (c and f, top) and $r_1/r_2{=}0.2$, $r_1/L{=}1.9$ (c and f, bottom) demonstrate that, for comparable $\widetilde{A}$, the confinement with higher eccentricity has more maximum density loops (c, highlighted in green), indicating high smectic order, as well as fewer minimum density loops (f, highlighted in green), indicating low tetratic order. Scale bars are 10 {\textmu}m.}
\label{fig:FigAAN-MinNetworkLoops}
\end{figure*}

Confinement of a liquid crystal not only controls defect formation but also shapes the energetic landscape of the system. With our rods having an aspect ratio ${\sim}8$, our system supports meta-stable, quasi-tetratic order, in addition to smectic order. By expanding the network analysis to identify local tetratic regions, we probe the effect of confinement geometry on meta-stable structures in a colloidal smectic. 

Local tetratic regions are surrounded on all sides by disclinations and as such have net-zero topological charge,  illustrated in Fig.~\ref{fig:FigAAK-MaxNetworkLoopsSchematic}. Charges in the density networks alone do not distinguish between tetratic regions and other defects that arise with smectic order. Instead, we look at closed loops in the network to quantify the degree of tetratic and smectic ordering. As shown in Fig.~\ref{fig:FigAAK-MaxNetworkLoopsSchematic}(a), closed maximum density loops capture smectic ordering by representing either two adjacent smectic layers that stretch between two boundaries or a smectic dislocation. Conversely, as shown in Fig.~\ref{fig:FigAAK-MaxNetworkLoopsSchematic}(b), closed minimum density loops capture tetratic ordering by enclosing a local region with rods differing in orientation by $\pi/2$ compared to their surroundings. We quantify the effect of confinement on smectic versus tetratic ordering by plotting the number of maximum and minimum density loops against the confinement area for varying eccentricity, shown in Fig.~\ref{fig:FigAAN-MinNetworkLoops}. 

Fig.~\ref{fig:FigAAN-MinNetworkLoops}(a) and (b) plot the number of loops in the maximum density network, $n_\text{max}$, for experiments and simulations, respectively. Confining geometries with high eccentricity, with $r_1/r_2=0.2$, have larger values of $n_\text{max}$ compared to more circular confinements, most notably at moderate system areas. A system with a moderate, normalized area $\widetilde{A} = 390$ is shown at the bottom of Fig.~\ref{fig:FigAAN-MinNetworkLoops}(c), where maximum density loops are highlighted in green. Higher $n_\text{max}$ values corroborate that high confinement eccentricity facilitates smectic ordering.

However, for larger systems, $n_\text{max}$ starts to scale linearly with the system area, due to the appearance of meta-stable, tetratic ordering in the bulk. Local, quasi-tetratic regions appearing in the center of larger systems is a consequence of an anchoring extrapolation length limiting the degree of order that a surface can impose into the bulk \cite{Kralj1996,Vaupotic1996,Kutnjak2004,Blanc2023}. The quasi-tetratic regions introduce disclinations that disrupt the formation of maximum density loops. This is seen in Fig.~\ref{fig:FigAAN-MinNetworkLoops}(a) and (b) for $r_1/r_2=0.2$, where $n_\text{max}$ approaches values seen in more circular systems with larger confinement area.  In more circular systems, disclinations are always formed due to the confinement shape promoting planar anchoring, disrupting the smectic ordering. An unconfined system also forms disclinations due to meta-stable, short-ranged, and quasi-tetratic ordering. Therefore, when elliptical systems with $r_1/r_2=0.2$ become large enough, the bulk has more disclinations from forming localized, tetratic regions. The anchoring from the elliptical confinement can only suppress meta-stable, tetratic ordering in the vicinity of the boundary, within a distance set by the anchoring extrapolation length.

The increase of quasi-tetratic regions for $r_1/r_2=0.2$ with larger system areas is also evident in the number of loops in the minimum density network, $n_\text{min}$, plotted in Fig.~\ref{fig:FigAAN-MinNetworkLoops}(d) and (e). Across all confinement geometries, $n_\text{min}$ scales positively with system area, which is again a consequence of the limitations of the anchoring extrapolation length. Upon increasing the system size, the liquid crystal has more area available for bulk behaviour and is able to exhibit meta-stable, tetratic structures.

Yet, confinements with higher eccentricity generally have lower $n_\text{min}$ values and thereby less tetratic ordering than more circular confinements, which is consistent with confinement ellipticity promoting smectic ordering. Fig.~\ref{fig:FigAAN-MinNetworkLoops}(f) highlights confinement geometries with $r_1/r_2$ of $0.2$ and $0.6$, with comparable areas, $\widetilde{A} = 390$ and 380, respectively. The more circular confinement (Fig.~\ref{fig:FigAAN-MinNetworkLoops}(f), top) has $n_\text{min} = 3$, reflecting three local, quasi-tetratic regions. The more elliptical confinement (Fig.~\ref{fig:FigAAN-MinNetworkLoops}(f), bottom) has no tetratic regions, with $n_\text{min} = 0$. The anchoring extrapolation length in more elliptical confinements covers more bulk area than for circular confinements, fostering smectic order.

The closed minimum density and closed maximum density loops demonstrate that confinements with high eccentricity can suppress meta-stable, tetratic ordering, with the inverse being true for more circular systems. The anchoring conditions formed by ellipses with high regions of curvature can be leveraged to curb meta-stable structures, yielding long-range order.
 
\section{\label{sec:Conclusion}Conclusion}

To sum up, we have shown with both experiments and simulations that boundary curvature directs the anchoring and defect state of colloidal smectics. We tuned the boundary curvature by varying the size and shape of elliptical confinements. The transition from planar to homeotropic anchoring occurs at a local, critical radius of curvature equal to or less than twice the rod length, $R^*_\text{curv}\approx2L$. From analyzing local and global $P_2$ and $P_4$ order parameters, we found that the anchoring condition set by boundary curvature dictates the disclination state of the system. We then employed a network analysis to distinguish disclinations and dislocations by examining topological charges and loops in the minimum density and maximum density networks. Smectic defect types can be selected by adjusting the boundary curvature: planar anchoring from confinements with only large curvatures promotes disclinations, while homeotropic anchoring from highly curved, elliptical confinements promotes dislocations. Within ellipses of intermediate sizes, metastable structures can be suppressed in systems where the anchoring extrapolation length covers a large portion of the bulk, yielding long-ranged, smectic order.

Interestingly, experiments and simulations agree well on the effect of boundary curvature on disclinations but have slight differences in the value of $R^*_\text{curv}$ and the number of dislocations. We attribute this to the polydispersity of rods in experiments that is absent in simulations. Disclinations carry an excess network charge of $\sum q =+1/2$ and are thereby greatly affected by the topological frustration from confinement. The topological frustration imposed on the system is unaffected by rod polydispersity. However, dislocations have a neutral topological contribution ($\sum q = 0$) and are thereby more easily influenced by local differences in microscopic interactions and macroscopic properties. Polydispersity in the rod shape and size alters the excluded-volume interactions of the rods with a hard-wall boundary, potentially impacting the rod anchoring. Polydispersity could also alter macroscopic, elastic properties. In their recent work connecting the microscopic interactions of hard rods to mesoscale elasticity \cite{Wensink2023}, Wensink and Grelet determined that the positional fluctuation of rods normal to their layer leads to a reduction in the layer bending modulus $K$ in Eq.~\ref{Eq:SmElasticity}. Polydispersity necessarily leads to rods with lengths that deviate from the average smectic layer size, and the length disparity of a rod has been found to increase its inter-layer diffusion \cite{Chiappini2020}. The smectic layers around a dislocation experience significant bend deformations, so polydispersity in rod length could plausibly lower the energy required to form a dislocation by lowering $K$. The effect of rod polydispersity on the mesoscale, elastic energy of lyotropic smectics remains to be fully elucidated and requires further study.

To conclude, we have demonstrated anchoring and defect control of hard-rods smectics using only hard-wall interactions. With the rod anchoring only dependent upon the local wall geometry, our findings are applicable to liquid crystalline systems across length scales. The conclusions from this work help to establish design principles for the self-assembly and defect control of larger-scaled liquid crystals, such as anisotropic cells \cite{Doostmohammadi2016,Kawaguchi2017,Saw2017,Maroudas-Sacks2021,VafaMahadevan2022,Kaiyrbekov2023} and functionalized nanorods \cite{Vroege2006,Querner2008,Zanella2011,Diroll2015,Hosseini2020,Hussain2021,Jehle2021} --- important for the development of bio- and nano-technologies. 

\begin{acknowledgments}
We thank Dave van den Heuvel, Relinde van Dijk-Moes, Albert Grau Carbonell, and Roy Hoitink for experimental support. We thank Amir Raoof for access to clean-room facilities. We also thank Alfons van Blaaderen, Arnout Imhof, and Randall D. Kamien for helpful discussions. E. I. L. J. acknowledges funding from the European Commission (Horizon-MSCA, Grant No. 101065631). G. C.-V. acknowledges financial support from the Netherlands Organization for Scientific Research (NWO) ENW PPS Fund 2018-Technology Area Soft Advanced Materials (Grant No. ENPPS.TA.018.002). M. D. acknowledges financial support from the European Research Council (Grant No.
ERC-2019-ADV-H2020 884902 SoftML). L. T. acknowledges support from the European Commission (Horizon-MSCA, Grant No. 101065631) and the NWO ENW Veni grant (Project No. VI.Veni.212.028). E. I. L. J. and L. T. acknowledge support from the Starting PI Fund for Electron Microscopy Access from Utrecht University's Electron Microscopy Center. 
\end{acknowledgments}

\pagebreak
\appendix

\section{Synthesis of fluorescent colloidal rods}\label{App:ColloidalSynthesis}

Rod-like silica colloids were synthesized following the method developed by Kuijk \textit{et al.} \cite{kuijk2011}. $30.1$ g polyvinylpyrrolidone (PVP, MW 40,000, Sigma-Aldrich) was fully dissolved via a combination of sonication and agitation in $300$ mL 1-pentanol (${\geq}99{\%}$, Honeywell). $30$ mL absolute ethanol (Baker), $10$ mL ultrapure water (Millipore system), and $2$ mL 0.19 M sodium citrate dihydrate (${\geq}99{\%}$, Sigma-Aldrich) in water was mixed with the PVP-pentanol solution. $6.75$ mL ammonium hydroxide (${\geq}25{\%}$, Sigma-Aldrich) was added, and the solution was shaken by hand to generate a micro-emulsion. The solution was left to stand for $120$ seconds to allow emulsion stabilization. A total of $12$ mL tetraethyl orthosilicate (TEOS, Sigma-Aldrich) was added to the solution in 4 steps of $3$ mL with 6 hour intervals. After each addition, the bottle was inverted to assist mixing of the TEOS without disrupting the emulsion. After the final TEOS addition, the solution was left for 24 hours to ensure completion of the reaction. The resulting colloids were cleaned with several rounds of centrifugation and supernatant removal, details of which can be found in Appendix \ref{App:ColloidClean}.

A fluorescent shell was then grown around the silica rods using the method developed by Kuijk \textit{et al}.~\cite{kuijk2014}. $9.5$ mg rhodamine-B-isothiocyanate (RITC, Sigma) dye was dissolved in $1.34$ mL absolute ethanol before adding 9.5 {\textmu}L (3-aminopropyl)triethoxysilane (APTES, ${\geq}98{\%}$ Sigma-Aldrich). The mixture was left overnight to react. Approximately 0.02 g of bare silica rods were isolated and dispersed in $100$ mL absolute ethanol. $5$ mL ammonium hydroxide solution and $5$ mL ultrapure water were added to the solution. 10.3 {\textmu}L TEOS and 53.3 {\textmu}L RITC-APTES mixture were added three times with 60 minute intervals between each addition (total 31 and 160 {\textmu}L TEOS and RITC-APTES mixture, respectively). This procedure resulted in a fluorescent shell of approximately $73{\pm}2$ nm. An additional silica shell was grown around the fluorescent shell by dispersion of the  RITC-labeled silica rods in $300$ mL absolute ethanol, $10$ mL ultrapure water, and $10$ mL ammonium hydroxide solution. 9.52 {\textmu}L TEOS were added three times with intervals of $60$ minutes for a total of 28.56 {\textmu}L, resulting in an additional silica shell of approximately $103{\pm}2$ nm. In both shell-growth steps, the solution was left over night to ensure the reaction was complete. 

\section{Cleaning synthesized rods}
\label{App:ColloidClean}
To remove particles produced in secondary nucleation during synthesis, the colloidal rods were cleaned by centrifugation. The suspensions were centrifuged with a force of 1500 g for 30 minutes to form a pellet, after which, the supernatant was discarded. This procedure was repeated 4 times. The suspension was then left to stand in 40 mL glass vials for 2 hours, collecting the top 2 mL of supernatant now populated with rods. The stock solution was then redispersed. This process was repeated until the desired rod concentration was obtained. The collected solution was then cleaned using centrifugation forces of 300, 150, 100, 75, and 50 g, each for 15 minutes. After each centrifugation step, the supernatant was discarded. Before moving to the next lower force, each centrifugation at a specific force was repeated until a clear supernatant was achieved. After each fluorescent and non-fluorescent silica shell growth procedure (see Appendix~\ref{App:ColloidalSynthesis}), the sample was cleaned at 700 g for 15 minutes, after which the supernatant was discarded.

\section{Photolithography of confinement wells}
\label{App:Photolith}
Micron-scale confinement was formed on ${\#}1$ Cover glass (130-170 {\textmu}m thickness, VWR) using standard contact-photolithography procedures. Briefly, the photopolymer SU-8 2010 (Micro Resist Technology) was spin-coated (WS-650-23B, Laurell Technologies Corporation) at 300 rpm for 10 seconds and then 1500 rpm for 30 seconds, to achieve an approximate film thickness of 10 {\textmu}m. The coverslip was then subject to edge bead removal, helping to minimise the air gap between the photomask and the photoresist. The coverslip was soft-baked at 95 $^{\circ}$C for 3 minutes. A chrome patterned photomask (JD PhotoData) was placed in contact with the substrate, and a UV exposure of 210 mJ/ cm$^2$ (30 mW/cm$^2$, at 30\% power) was conducted with UV-EXP150R UV exposure system (idonus s\`{a}rl). A subsequent post-exposure bake at 95 $^{\circ}$C for 4 minutes was done prior to submerging the coverslip in developer (mr-Dev 600, Micro Resist Technology) for 10 minutes. The coverslips were then rinsed with isopropyl alcohol before a hard bake at 200 $^{\circ}$C for 60 minutes, to thermally set the remaining polymer.

\section{Optical characterization}
The synthesized silica rods were characterized using transmission electron microscopy (Thermo Fisher Scientific Tecnai F20) with an acceleration voltage of 200 kV. Samples were prepared by drop casting the rod particles dispersed in ethanol onto Formvar/Carbon 200 mesh copper grids ($\#$01801, Ted Pella).

The assembled, fluorescently-labeled silica rods were imaged with confocal fluorescent microscopy, using an inverted Leica TCS SP8 STED 3X confocal microscope. A 63x/1.30 glycerol objective was used with an immersion liquid of 85 wt\% glycerol in water. The fluorescent dye RITC that was used to label the silica rods was excited using a pulsed (80 MHz) super-continuum white light laser (SuperK, NKT Photonics), which was tuned to 543 nm. The emission was then detected using a gated (0.3-6.0 ns) HyD detector with a detection window of 554 to 691 nm.

\section{Bulk behavior}
\label{App:Bulk behavior, Bates and Frenkel}
\begin{figure}
\includegraphics[width=0.45\textwidth]{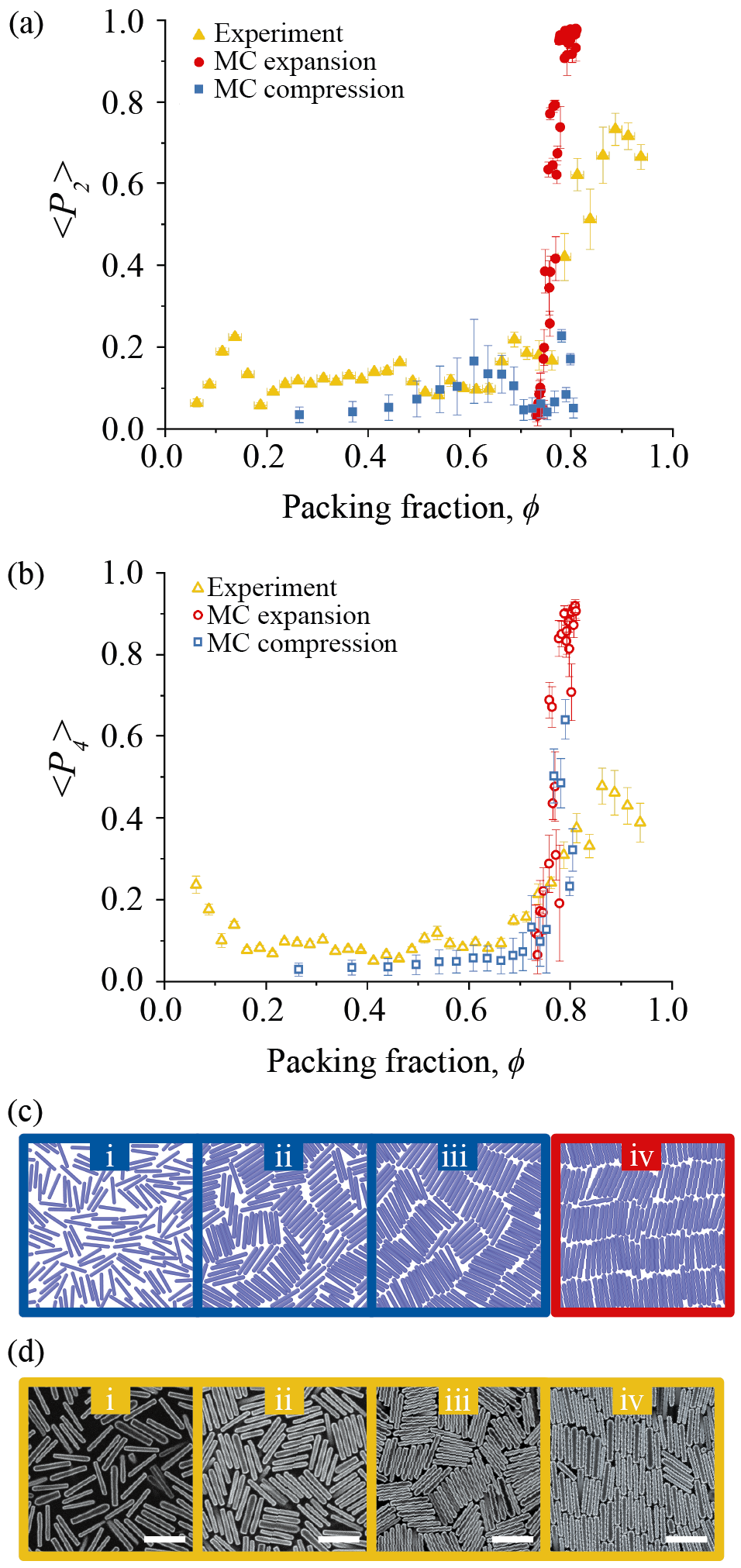}
\caption{(a) Nematic $\left<P_2\right>$ and (b) tetratic $\left<P_4\right>$ order parameter as a function of packing fraction $\phi$ for MC compression simulations (blue squares), MC expansion simulations (red circles), and sedimentation-diffusion experiments (yellow triangles). A phase transition from isotropic to smectic/tetratic is observed at packing fractions of $\phi \sim 0.75$. (c) Snapshots of MC compression (i-iii) and expansion (iv) simulation snapshots at packing fractions of 0.44, 0.72, 0.79, and 0.80, respectively from i-iv. (d) Snapshots of experiments at packing fractions of 0.49, 0.72, 0.8, and 0.8 (after additional relaxation time), from i-iv, demonstrating, with increasing $\phi$, the phase transition from isotropic to smectic with intermediate tetratic ordering.}
\label{fig:AAB-UnconfinedData}
\end{figure}

In line with predictions from Bates and Frenkel~\cite{Frenkel2000}, the smectic phase is not accessible under compression, demonstrated by the low $\left<P_2\right>$ values with increasing packing fraction $\phi$ (blue squares, Fig.~\ref{fig:AAB-UnconfinedData}(a)). Instead, a (meta-stable) phase characterized with short-range tetratic order is formed (Fig.~\ref{fig:AAB-UnconfinedData} (c, iii)), evidenced by an increase in $\left<P_4\right>$ at ${\phi}{=}0.75$ (open-blue squares, Fig.~\ref{fig:AAB-UnconfinedData}(b)). Again similar to the findings of Bates and Frenkel, a clear transition from smectic (Fig.~\ref{fig:AAB-UnconfinedData}(c, iv.)) to isotropic is observed at ${\phi}{=}0.75$ under expansion, demonstrated by the sharp decrease in $\left<P_2\right>$ and $\left<P_4\right>$ with reducing $\phi$ (red circles, Fig.~\ref{fig:AAB-UnconfinedData}(a)).

Sedimentation experiments of silica rods behaves effectively like a slow compression with a gradual increase in packing fraction. At packing fractions $\phi\leq 0.75$ both the nematic and tetratic order parameters remain at isotropic values of ${\sim}0.1$. At packing fractions $\phi {\geq} 0.75$ a phase transition is observed, with an increase in both $\left<P_2\right>$ and $\left<P_4\right>$ to approximately $0.7$ and $0.4$, respectively (yellow and open-yellow triangles in Fig.~\ref{fig:AAB-UnconfinedData}(a) and (b)). In experiments, the tetratic ordering can be relaxed away when the system is allowed to equilibrate for long times ($\sim5$ months). We hypothesize that flow effects, rod polydispersity, and out-of-plane rod fluctuations enable structural relaxations towards an well-ordered smectic phase (Fig.~\ref{fig:AAB-UnconfinedData} (d, iv)). Even at high $\phi$ where the ordering is predominantly smectic, localized clusters of tetratic ordering can be observed (Fig.~\ref{fig:AAB-UnconfinedData}(d, iii) and (d, iv)). 

There is good agreement between experimental and simulated phase behavior for rods with end-to-end, length-to-diameter ratios of ${\sim}8$. An isotropic-to-smectic phase transition is observed at $\phi\approx0.75$ for both simulations and experiments. The phase diagram demonstrates intermediate long-lived tetratic ordering in MC compression simulations, which appears to be meta-stable in experimental sedimentation-diffusion measurements. 

The experimental system can likely access the smectic phase because each rod is capable of moving into the third dimension. The gravitational length of the silica rods is on average 56 nm, which is much smaller than the rod diameter. Yet, we observe in localized areas of the samples that rods can be pushed out of the plane by neighboring rods (see Supporting Video 1). We hypothesize that each rod that is pushed out of the plane frees up area and behaves as a ``local expansion," allowing for the metastable, tetratic ordering to relax away. We note that confinements with high eccentricity act to further stabilize the smectic phase by providing a favored orientation direction for the global director (see Fig.~\ref{fig:AAC-GlobalP2}). 

\section{Measuring anchoring from snapshots}
\label{App:RcurvCalc}

To determine how the local curvature of a boundary affects the rod anchoring (see Fig.~\ref{fig:AAE-CurvatureAngleDevi}), we must determine the following quantities from experimental and simulated snapshots: 1) local radius of curvature $R_\text{curv}$ at a given point on the boundary, 2) the tangent line at that given point, and 3) average rod orientation near that given point. To calculate $R_\text{curv}$, we use Eq.~\ref{eq:EllipseRcurv}. The tangential angle $\varphi_\text{tangent}$ at the boundary, with respect to the x-axis, can be calculated in terms of the polar angle $t$ with $\varphi_\text{tangent}=\tan^{-1}(\frac{r_1}{-r_2}\frac{\cos{t}}{\sin{t}})$.

\begin{figure}
\includegraphics[width=0.48\textwidth]{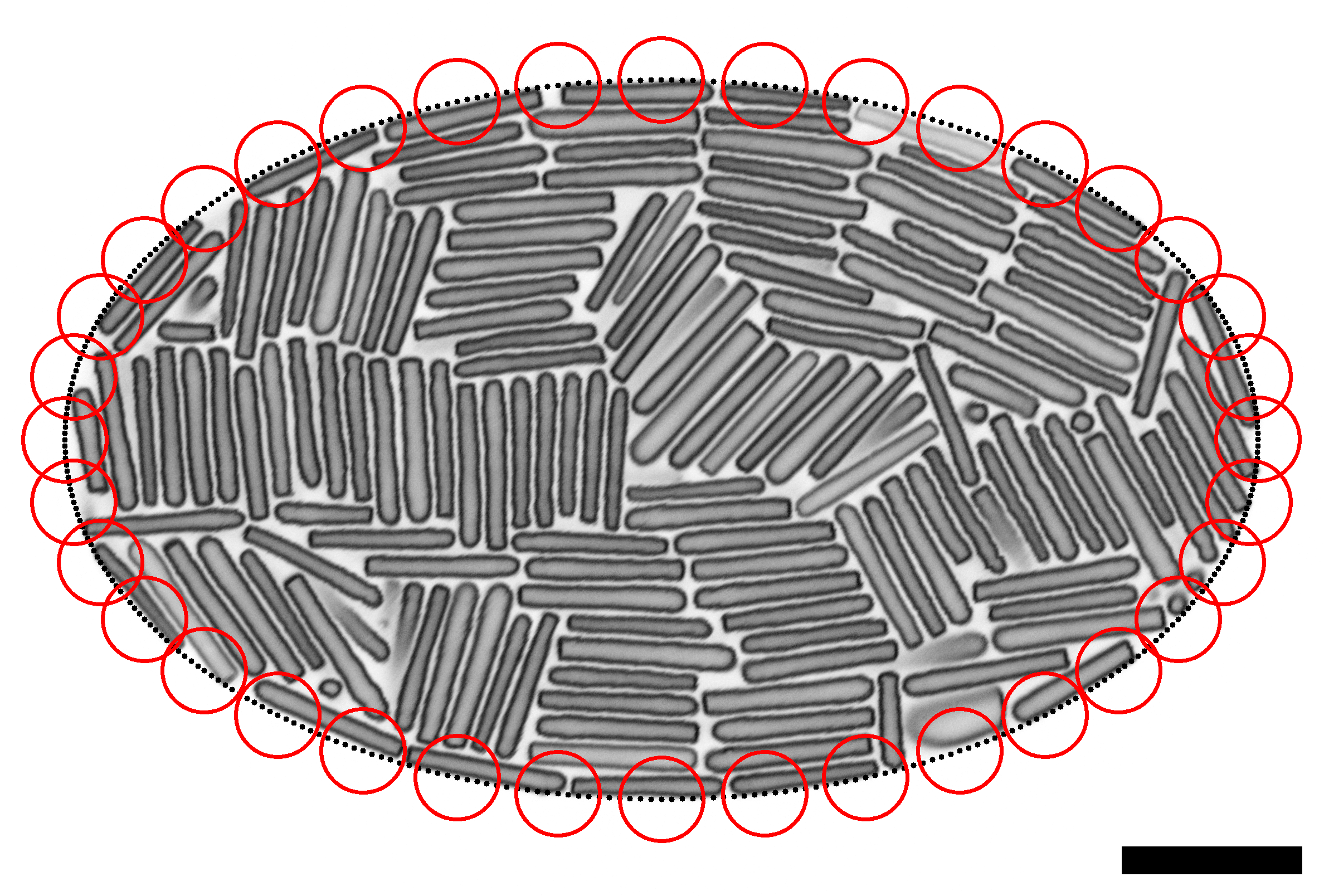}%Single column figure
\caption{Example colloidal smectic in an elliptical confinement with $r_1/r_2=0.61$, $r_1/L=2.4$. The scale bar is 10 {\textmu}m. The anchoring around the boundary is sampled by drawing a circle with radius $2D$ and determining the average rod orientation enclosed by the circle, weighted by the area of the rod within the circle. The anchoring around the entire boundary is examined in this fashion. We illustrate here the anchoring sampling with steps of 10 degrees in the polar angle $t$.}
\label{fig:Fig_BoundaryAnalysis}
\end{figure}

With parameters for a given point at the ellipse boundary defined, we can now examine the rod orientation around the entire ellipse boundary, as illustrated in Fig.~\ref{fig:Fig_BoundaryAnalysis}. To sample the rod orientation near a given point on the boundary, we take a circle with a radius of twice the rod diameter $2D$ around the point. The average rod orientation $\left<\varphi_\text{particle}\right>$ is determined and is weighted by the area of each rod enclosed within the circle. Then, we calculate the difference between $\left<\varphi_\text{particle}\right>$ and $\varphi_\text{tangent}$ to get the angle deviation $\sigma$ at a boundary point: $\sigma = |( \left<\varphi_\text{particle}\right> - \varphi_\text{tangent} )|$. The sampling around the boundary is done at intervals of $\Delta{t}=1^{\circ}$ for both experimental and simulated snapshots. We repeat this process around the ellipse boundary for systems with $r_2/L\geq2$, to exclude overly confined systems. A moving average with a step size of $0.05$ and a window size of $0.25$ is done for both data sets. The averaged values produce the data displayed in Fig.~\ref{fig:AAE-CurvatureAngleDevi}. 

\section{Network generation protocol}
\label{App:Network Generation Protocol}
\begin{figure}
\includegraphics[width=0.48\textwidth]{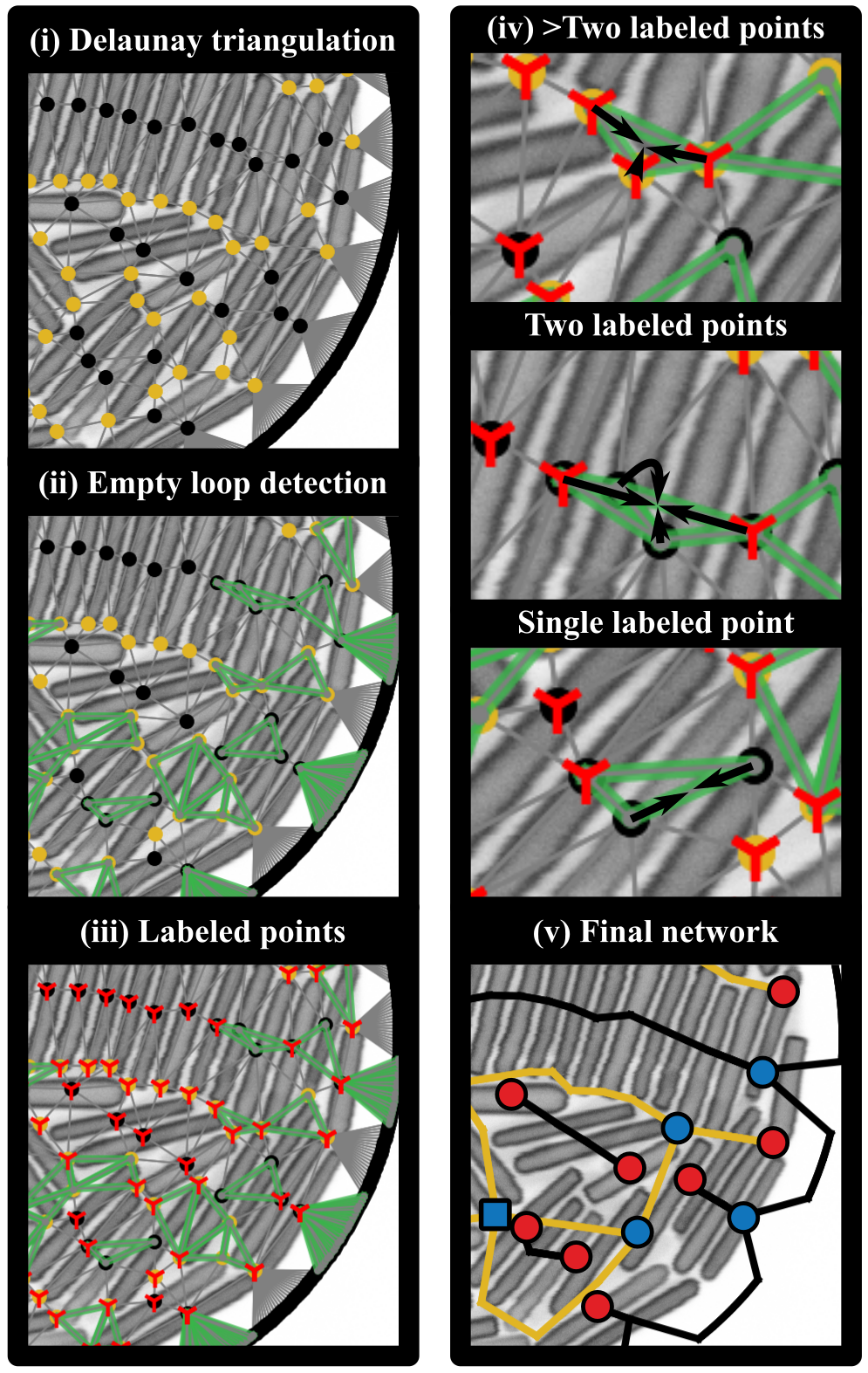}
\caption{The protocol for drawing minimum and maximum density networks starts with the (i) Delaunay triangulation. (ii) Empty loops are then detected in each network species, highlighted in green. (ii) Hierarchically important vertex points are detected, labeled by red Y's. (iv) Vertex merging is dependent upon the number of labeled points contained in the empty loop. From top to bottom, the protocol is visually represented for more than two fixed points, two fixed points, and a single fixed point. Vertex merging is illustrated by black arrows. (v) The final network with associated topological charge is shown.}
\label{fig:SI-B-NetworkAnalysis}
\end{figure}

The network analysis of a particle-resolved, colloidal smectic starts with the definition of vertex points corresponding to the particle center of mass (maximum density network) and particle ends (minimum density network). To prevent excessive empty loop generation, due to neighboring rod ends, the minimum density network points are run through an averaging protocol which merges any minimum density vertices to a single point that are within a single rod diameter of one another. Vertex points for the elliptical boundary are labeled and assigned to the maximum density network to ensure topological conservation. A Delaunay triangulation is then done for all points in the network, as shown in Fig.~\ref{fig:SI-B-NetworkAnalysis}(i).

The two networks are then separated from one another by removing connections between vertex points of opposing networks. Next, to ensure that the final networks represent the expected smectic layer structure, empty loops are collapsed to simple lines or points, where an empty loop is defined as a triangular mesh made up of a single network species that does not enclose any portion of the opposing network. The collapsing of empty loops must be done carefully to ensure that there are no hierarchical changes between the two networks. In order to merge the empty loops without changing the network hierarchy, we first identify the empty loops [Fig.~\ref{fig:SI-B-NetworkAnalysis}(ii)] and then label vertex points that are important for maintaining the network hierarchically (Fig.~\ref{fig:SI-B-NetworkAnalysis}(iii)). 

More specifically, empty singular loop detection starts with the identification of empty triangle meshes composed of a single network species. Empty loops made up of neighboring singular loops are then detected. If two empty triangular meshes, of the same network species, share an edge, then they are considered to be part of the same ``macroscopic" empty loop [see green highlighted loops in Fig.~\ref{fig:SI-B-NetworkAnalysis}(ii)]. 

Next, vertex points are labeled if they fulfill any of the following criteria and are shown by red Y's in Fig.~\ref{fig:SI-B-NetworkAnalysis}(iii):
\begin{itemize}
    \item A vertex point of either network species is connected to the boundary.
    \item Two connected vertices of the same network species are not contained within an empty loop.
    \item A vertex of a given network species is the single connection point between two separate empty loops of the same species.
\end{itemize}

Now with empty loops and labeled vertices detected, the empty loops are merged into a single vertex following a specific order in the procedure, listed below. The resultant vertex is determined according to the number of labeled vertices contained within the empty loop boundary, as follows:
\begin{enumerate}
    \item Greater than two labeled points --- \textit{all} points in the empty loop are merged to the average vertex point of \textit{all} points (Fig.~\ref{fig:SI-B-NetworkAnalysis}(iv), top).
    \item Two labeled points --- \textit{all} points in the empty loop are merged to the average vertex point of the two labeled points (Fig.~\ref{fig:SI-B-NetworkAnalysis}(iv), middle).
    \item Single labeled point --- \textit{non-fixed} points in the empty loop are merged to the average vertex point of the \textit{non-labeled} vertices (Fig.~\ref{fig:SI-B-NetworkAnalysis}(iv), bottom).
\end{enumerate}

Once the empty loops are merged accordingly (see Fig.~\ref{fig:SI-B-NetworkAnalysis}(iv) for examples), the final networks can be plotted, and topological charge can be assigned to each network. Importantly, interstitials are manually removed from the analysis, and dislocation defects are checked to ensure correct negative topological charge allocation to the maximum density network (see Fig.~\ref{fig:FigAAJ-NetworkAnalysisDefects}).
\nocite{*}

\section{Effect of rod polydispersity}\label{app:Polydispersity}

\begin{figure*}
\includegraphics[width=0.92\textwidth]{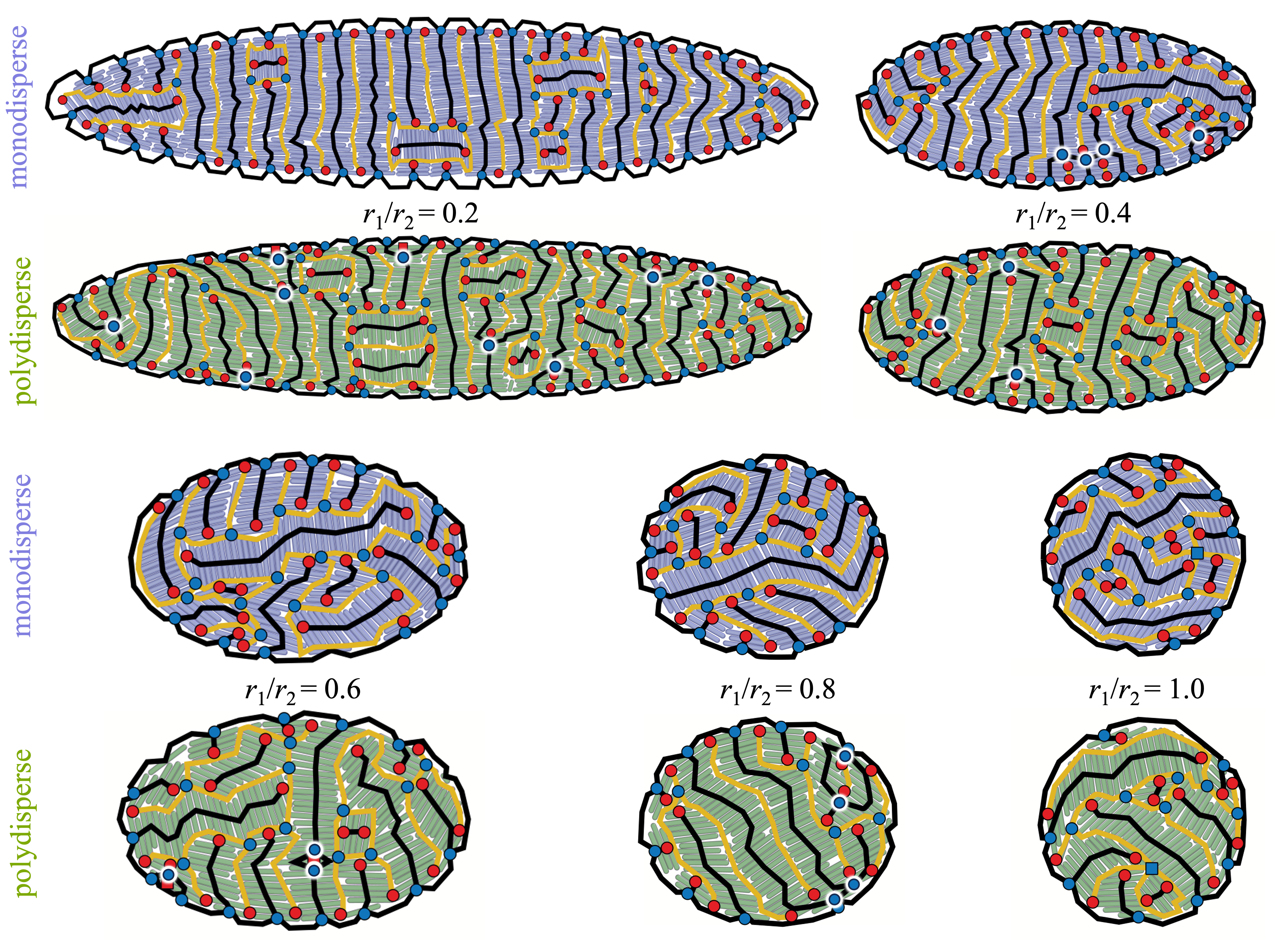}
\caption{Simulation snapshots of colloidal smectic rods with monodisperse (purple) and polydisperse (green) rod lengths confined within ellipses with varying $r_1/r_2$ and with the dimension set by $r_1/(7D)=2.55$. The minimum and maximum density networks are drawn in yellow and black lines, respectively. Positive and negative charges on the networks are labeled by red and blue dots, respectively. Internal negative charges on the maximum density network corresponds to dislocations and are highlighted by a white border. Polydispersity tends to increase the amount of dislocations, compared to monodisperse systems.}
\label{fig:polydispersity}
\end{figure*}

We conduct further MC simulations on polydiserse spherocylinder systems to explore their topological characteristics. Specifically, our goal is to illustrate that polydispersity predominantly drives the emergence of dislocations (see Fig~\ref{fig:polydispersity}). The experimental samples contain particles with (narrow) distributions in both length and diameter. However, for simplicity, in our simulations we match the experimental polydispersity in the end-to-end, length-to-diameter ratio of the rods ($L/D$) keeping $D$ equal for all particles. As in the case of the monodisperse system, we use the algorithm of Vega and Lago~\cite{Vega1994} to determine the shortest distance between rods and thus whether or not they overlap. 

We first verify that a polydisperse system of $N=1000$ particles, with $N_r=10$ different length-to-diameter ratios (or equivalently, particle types), does not undergo a demixing transition and that isotropic and smectic-like phases are stable. Then, in order to make a comparison with the monodisperse case, we consider systems confined within ellipses of different shapes by varying $r_1/r_2$ and with dimension set by $r_1/(7D)=2.55$. As in the monodisperse systems, we set the packing fraction at $\phi = \sum_{i=1}^{N_r}\phi_{i}=(1/A)\sum_{i=1}^{N_r}N_{i}A_{p,i}=0.81$, where $A_{p,i}=(L_{i}-D)D + \pi D^{2}/4$ and $N_{i}$ are the area of particle $i$ and the corresponding number of particles of type $i$, respectively. In simulations of these polydisperse systems, we are unable to implement cluster rotation moves and therefore, simulations are typically run for up to $4 \times 10^{7}$ steps.

\begin{table*}
\begin{tabular}{@{}lrrrrrrrrrr@{}}
\toprule
$r_1/r_2$ & \multicolumn{2}{c}{0.2}  & \multicolumn{2}{c}{0.4} & \multicolumn{2}{c}{0.6}  & \multicolumn{2}{c}{0.8}  & \multicolumn{2}{c}{1.0} \\ 
\midrule
$\sum q^-_\text{max}$ & {\color[HTML]{817FDF} 0}  & {\color[HTML]{7EAA05} -4.5}  & {\color[HTML]{817FDF} -2} & {\color[HTML]{7EAA05} -1.5} & {\color[HTML]{817FDF} 0}   & {\color[HTML]{7EAA05} -1.5} & {\color[HTML]{817FDF} 0}    & {\color[HTML]{7EAA05} -2}   & {\color[HTML]{817FDF} 0}  & {\color[HTML]{7EAA05} 0}    \\ 
\midrule
$\sum q^+_\text{max}$ & {\color[HTML]{817FDF} +15.5} & {\color[HTML]{7EAA05} +16.5} & {\color[HTML]{817FDF} +8} & {\color[HTML]{7EAA05} +9.5} & {\color[HTML]{817FDF} +10} & {\color[HTML]{7EAA05} +7.5} & {\color[HTML]{817FDF} +6.5} & {\color[HTML]{7EAA05} +3.5} & {\color[HTML]{817FDF} +7} & {\color[HTML]{7EAA05} +5.5} \\ 
\bottomrule
\end{tabular}
\label{tab:polydispersity}
\caption{Negative ($\sum q ^-_\text{max}$) and positive ($\sum q ^-_\text{max}$) internal charges on the maximum density network for the confinements shown in Fig.~\ref{fig:polydispersity}. For each confinement shape set by $r_1/r_2$, network charge values for monodisperse systems are given in the left column (purple), and values for polydisperse systems are in the right column (green). The increase in $\sum q ^-_\text{max}$ indicates that more dislocations arise with polydispersity in the rod length.}
\end{table*}

Polydispersity generally increases the number of negative charges in the maximum density network, as shown in Fig.~\ref{fig:polydispersity} and Table~\ref{tab:polydispersity}. For $r_1/r_2 =$ 0.2, 0.6, and 0.8, the monodisperse systems have no dislocations with $\sum q^-_\text{max} = 0$, but polydisperse systems have $\sum q^-_\text{max} < 0$. For the most circular confinement with $r_1/r_2 = 1.0$, both monodisperse and polydisperse systems have no dislocations, as dislocations are suppressed by the small confinement size and planar anchoring. For $r_1/r_2 = 0.4$, the monodisperse system has a few dislocations, unlike the other confinement geometries. Therefore, the polydisperse system continues to have dislocations as well. The exact number of dislocations often fluctuates for a given confinement. Overall, the introduction of length and width polydispersity to the rods appears to promote the formation of dislocations. We conclude from this that the presence of polydispersity in experiments also results in more dislocations appearing in the experimental system compared to the monodisperse simulations.

\section{Smectic layer spacing}\label{App:SmecticLayerSpacing}

\begin{figure}
\includegraphics[width=0.48\textwidth]{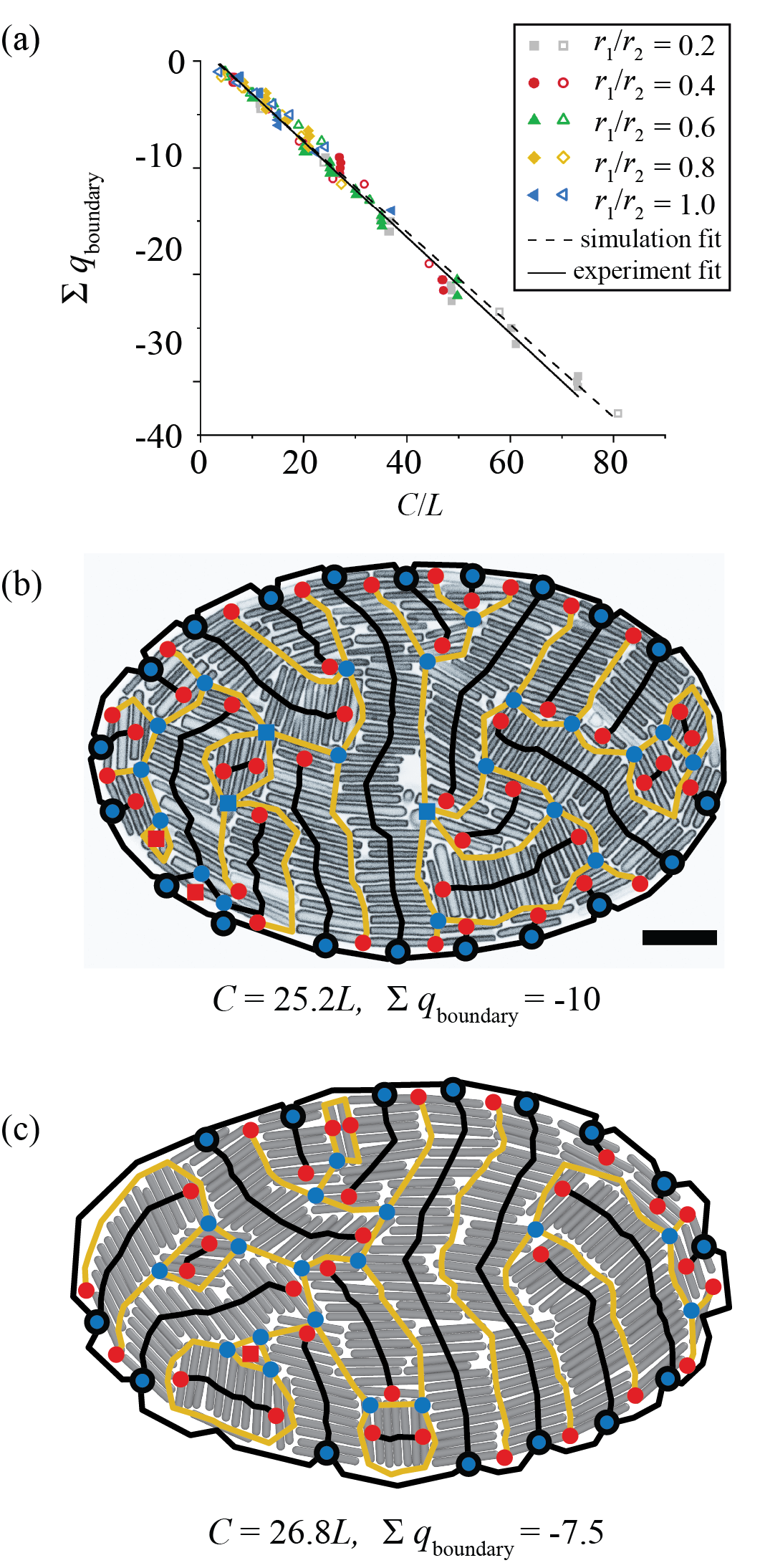}%Single column figure
\caption{(a) Total topological charge formed at the confining boundary as a function of the boundary circumference (C) normalized by the rod length ($L$). The experimental results (filled points) have a gradient of $-0.449{\pm}0.004$, while the simulated results (open points) have a lower magnitude gradient of $-0.432{\pm}0.009$. This indicates an increased smectic-layer spacing or increased amount of homeotropic anchoring away from regions of highest curvature in the simulated system. The increased homeotropic anchoring is demonstrated when comparing (b) experimental and (c) simulated snapshots for $r_1/r_2{=}0.6$, $r_1/L{\sim}3.1$. The simulated system has fewer boundary charges (vertices outlined in black) as a result of increased homeotropic anchoring. Experimental scale bar is 10 {\textmu}m.}
\label{fig:FigAAH-BoundaryDefects}
\end{figure}

To back out the smectic layer spacing from the network analysis, we can look at the topological charge associated with the boundary of each confining geometry and relate this to the confinement circumference (Fig.~\ref{fig:FigAAH-BoundaryDefects}). By assigning the boundary to the maximum density network, a rod that observes planar anchoring at the boundary creates a ${-}1/2$ topological charge on the boundary network. Since each rod at the boundary should be part of a smectic layer, an increase in the number of smectic layers results in additional ${-}1/2$ charges. Twice the magnitude of the charges on the boundary $|\sum q_\text{boundary}|$ is then the total number of smectic layers in the system. 

Fig.~\ref{fig:FigAAH-BoundaryDefects}(a) plots the boundary charge $\sum q_\text{boundary}$ against the confinement circumference $C$, normalized by the rod length $L$. Experimental data is plotted with solid points, while simulated data is plotted with open points. The data sets can be fit to a line, and the slope of the line can be used to calculate the smectic layer size. The magnitude of the slope of the fitted line for experiments (solid black) is approximately 0.45, while the magnitude of the slope for simulations (dotted black) is approximately 0.43. Multiplying these values by two and inverting it yields smectic layer sizes for experiments and simulations of ${\lambda}/L{=}1.11$ and  ${\lambda}/L{=}1.16$, respectively.

Independent smectic-layer spacing measurements from snapshots of \textit{unconfined} systems yield lower values, with ${\lambda}/L\approx1.08$ and ${\lambda}/L\approx1.05$ for the experimental and simulated systems, respectively. Comparing these values to those obtained from network analysis, we find that the network analysis slightly overestimates the smectic layer spacing. We attribute the measured increase in the smectic layer spacing under confinement to the ability of the rods to have homeotropic anchoring at the boundary, beyond the regions of highest curvature. Fig.~\ref{fig:FigAAH-BoundaryDefects}(b), on the right side of the experimental system, and Fig.~\ref{fig:FigAAH-BoundaryDefects}(c), on the left side of the simulated system, depict such regions of non-tangent anchoring. The increased amount of non-tangent anchoring results in fewer boundary charges, which increases the calculated smectic layer size. We tend to see more non-tangent anchoring away from the ellipse vertices for more circular confinements compared to more elliptical systems, as evidenced by the distribution of data points above or below the linear fits in Fig.~\ref{fig:FigAAH-BoundaryDefects}(a). More circular confinements with $r_1/r_2 \geq 0.8$ tend to sit above the fitted line, while more elliptical confinements with $r_1/r_2 = 0.2$ tend to sit below it.

Analysis of the boundary defects from the maximum density network allows for an approximation of the layer spacing for smectic colloidal liquid crystals under confinement. Increases in non-tangent anchoring at the boundary that extends beyond the regions of highest curvature leads to a slight overestimation of the value.

\newpage
%\bibliography{SmecticPaper}% Produces the bibliography via BibTeX.

\providecommand{\noopsort}[1]{}\providecommand{\singleletter}[1]{#1}%

\end{document}